RANDOM NUMBER GENERATOR, ZERO-CROSSING, AND NONLINEARITY ATTACKS AGAINST THE KIRCHHOFF-LAW-JOHNSON-NOISE (KLJN) SECURE KEY EXCHANGE PROTOCOL

A Dissertation

by

CHRISTIANA SÖKELAND FREITAS CHAMON

Submitted to the Office of Graduate and Professional Studies of
Texas A&M University
in partial fulfillment of the requirements for the degree of

DOCTOR OF PHILOSOPHY

| | |
|---|---|
| Chair of Committee, | Laszlo Kish |
| Committee Members, | Jun Zou |
| | Chanan Singh |
| | Andreas Klappenecker |
| Head of Department, | Miroslav Begovic |

May 2022

Major Subject: Electrical Engineering

# ABSTRACT

This dissertation demonstrates three new types of attacks against the Kirchhoff-Law-Johnson-Noise (KLJN) secure key exchanger.

The first attack type is based on compromised random number generators.

The first RNG attacks are deterministic. In the first attack, Eve knows both noises. We show that Eve can quickly crack the bit via Ohm's Law and one-bit powers, within a fraction of the bit exchange period. In the second attack, Eve knows only Bob's noise, so she can learn Bob's resistance value via Ohm's Law and Alice's resistance at the end of the bit exchange period. She can also use a process of elimination.

The second RNG attacks are statistical. In the first attack, Eve has partial knowledge of Alice's and Bob's noises. We show that Eve can crack the bit by taking the highest cross-correlation between her noises and the measured noise in the wire, and by taking the highest cross-correlation between her noises and her evaluation of Alice's/Bob's noises. In the second attack, Eve has partial knowledge of only Alice's noise. In this situation, Eve can still crack the bit, but after the bit exchange period.

The second attack type is based on thermodynamics. Previously, the KLJN scheme required thermal equilibrium. However, Vadai, et al, in (Nature) Science Reports shows a modified scheme, where there is a non-zero thermal noise, yet the system resists all the known attacks. We introduce a new attack against their system, which utilizes coincidence events between the line current and voltage. We show that there is non-zero information leak toward the Eavesdropper. As soon as the thermal equilibrium is restored, the system becomes perfectly secure again.

The final attack type is based on the nonlinearity of the noise generators. We explore the effect of total distortion at the second order, third order, and a combination of the second and third orders on the security of the KLJN scheme. It is demonstrated that a distortion as little as 1% results in a notable power flow, which leads to a significant information leak. We also show that decreasing the effective temperature results in the KLJN scheme approaching perfect security.

# DEDICATION

To Theo, my heart, my sunshine.

# ACKNOWLEDGMENTS

First and foremost, I would like to thank my advisor Dr. Laszlo Bela Kish for taking me in as a "refugee" Ph.D. student and for his continual friendship, mentorship, guidance, humor, and support throughout the remainder of my time here at Texas A&M University and for providing me the training wheels needed to live up to my potential as a scientist. Switching to his research group is literally the best decision I've ever made as a graduate student. I would also like to thank my committee members Dr. Jun Zou, Dr. Chanan Singh, and Dr. Andreas Klappenecker for being here for me and for their friendly encounters.

I would also like to thank Dr. Jose Silva-Martinez, Dr. Edgar Sanchez-Sinencio, and Dr. Kamran Entesari for being such great professors and for the friendly encounters, conversations, and laughs, as well as those from Ms. Ella Gallagher. I always enjoyed their company, and I would especially like to thank Dr. Sanchez for keeping the coffee machine constantly running! The graduate office crew Melissa Sheldon, Katie Bryan, and Dr. Scott Miller also have my thanks for their academic advising and for the aforementioned.

My main collaborator Shahriar Ferdous has my thanks for actively working with me and for his never-ending friendship. My thanks also extends to my friends in the ECEN department for their friendship, mentorship, and keeping me happy, as well as my friends outside of ECEN and Texas A&M University. A special thanks goes to my friend and former professor Dr. Gülin Tulunay Aksu from the University of Houston for being here for me, keeping me grounded in reality, and continually providing evaluations and suggestions in and out of school from the perspective of a former graduate student.

My gratitude further extends to my AggieWesties family who makes me look like a good West Coast Swing dancer every Wednesday night as I escape from reality. It was truly an honor to serve as the treasurer and vice president of the greatest organization in the world. The same thanks also extends to my Westie friends outside of AggieWesties. My Dance Barre family also holds a special place in my heart, and I cannot thank them enough for helping me improve my solo technique, continually pushing me to be the best dancer I can be, providing me a safe space, and for being here for me throughout the COVID-19 pandemic when the dance floor was forcibly taken away from me.

I am grateful for my parents Jorge and Ana Chamon, my siblings Julia and Gabriel Chamon, my parents-in-law Lusi and Doroteo Garcia, and my siblings-in-law Michael and Bianca Ortega and Juan and Jasmine Garcia for their constant care and support, even though I am 100 miles away. This thanks also extends to my prickle, for continually including me in their lives and cheering for me no matter how far away they are or how busy their lives got.

Penultimately but far from least, I am forever indebted to my wonderful loving husband, biggest fan, and best friend in the world Doroteo "Theo" Garcia for moving to College Station to be with me, for his exceeding patience, for constantly supporting and cheering for me, and for believing in me even when I didn't believe in myself. Being married to a graduate student is hard. However, he pulled it off so amazingly. It is to him that I dedicate this dissertation to.

My Ph.D. studies would not have been possible without these people, for which I stand eternally grateful. Finally, my eternal gratitude extends to God, for it is He who has ultimately provided me with this opportunity ("Jesus looked at them and said, 'With man this is impossible, but with God all things are possible.'" ~Matthew 19:26).

# CONTRIBUTORS AND FUNDING SOURCES

**Contributors**

This work was supported by a dissertation committee consisting of Dr. Laszlo Kish [advisor], Dr. Jun Zou, and Dr. Chanan Singh of the Department of Electrical Engineering and Dr. Andreas Klappenecker of the Department of Computer Science.

The Gaussian band-limited white noise was provided by Shahriar Ferdous.

All other work conducted for the dissertation was completed by the student independently.

**Funding Sources**

Graduate study was supported by a Teaching Assistantship, the Ebensberger Fellowship, and grader positions, all from Texas A&M University.

# NOMENCLATURE

| | |
|---|---|
| BEP | Bit Exchange Period |
| D | Distortion |
| FCK | Ferdous, Chamon, and Kish |
| FFT | Fast Fourier Transform |
| GBLWN | Gaussian Band-Limited White Noise |
| IFFT | Inverse Fast Fourier Transform |
| KLJN | Kirchhoff-Law-Johnson-Noise |
| NSA | National Security Agency |
| NSS | National Security Systems |
| QKD | Quantum Key Distribution |
| RMS | Root Mean Square |
| RNG | Random Number Generator |
| TD | Total Distortion |
| VMG | Vadai, Mingesz, and Gingl |

TABLE OF CONTENTS

LIST OF FIGURES

LIST OF TABLES

# 1. INTRODUCTION[1,2]

## 1.1 On Secure Communications

One way to establish the security of a communication is through encryption, that is, the conversion of plaintext into ciphertext via a cipher [1]. Figure 1.1 provides the general scope of symmetric-key cryptography [1]. The key is a string of random bits and both communicating parties Alice and Bob use the same key and ciphers to encrypt and decrypt their plaintext.

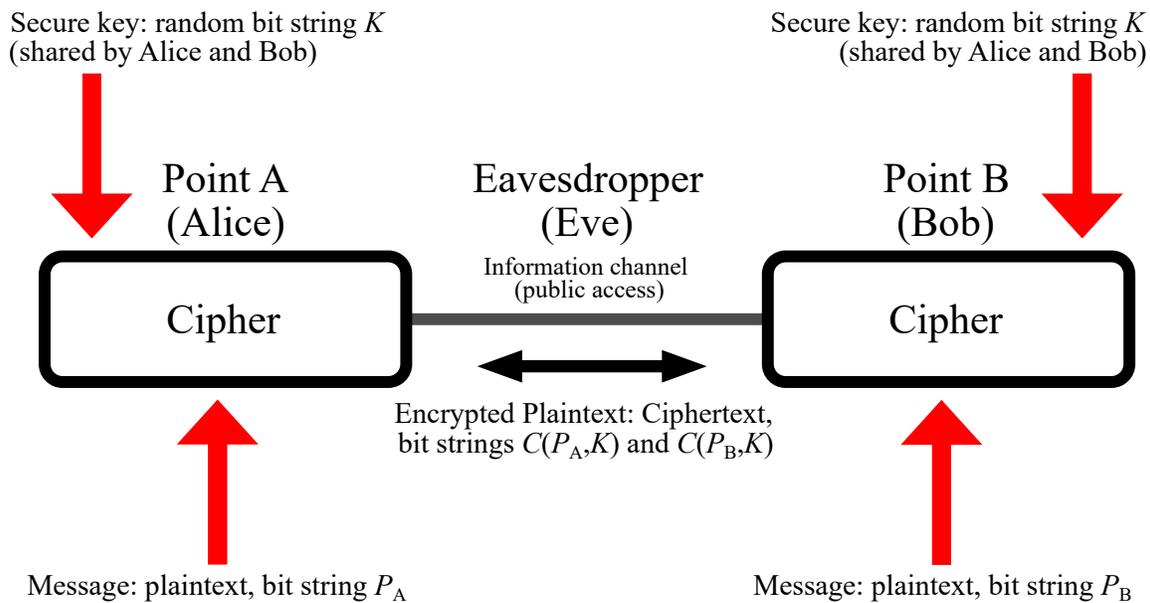

Figure 1.1: Symmetric-key cryptography [1]. Alice and Bob securely exchange a key (a string of random bits) through an information channel. The ciphers encrypt plaintext into the ciphertext $C$. The secure key is $K$, the plaintext messages of Alice and Bob are denoted by $P_A$ and $P_B$, respectively, and the ciphertext is a function $C(P,K)$.

For a plaintext message $P$ and a secure key $K$, the encrypted message, or the ciphertext $C$, is a function of $P$ and $K$, that is,

$$C = C(P, K). \tag{1.1}$$

In symmetric-key cryptography, for decryption, the inverse operation is used:

$$P = C^{-1}[C(P, K), K]. \tag{1.2}$$

Because the secure keys must be the same at the two sides (shared secret), another type of secure data exchange is needed before the encryption can begin: the secure key exchange, which is the generation and distribution of the secure key over the communication channel. Usually, this is the most demanding process in the secure communication because the communication channel is accessible by Eve, thus the secure key exchange is itself a secure communication where the cipher scheme shown in Figure 1.1 cannot be used. Eve records the whole communication during the key exchange, too. She knows every detail of the devices, protocols, and algorithms in the permanent communication system (as stated by Kerckhoffs's[3] principle/Shannon's maxim [2]), except for the key. In the ideal case of perfect security, the key is securely generated/shared, immediately used by a One Time Pad [3], and discarded after the usage. In practical cases, usually there are deviations from these strict conditions, yet the general rule holds: a secure system cannot be more secure than its key.

Our focus topic is the unconditionally secure Kirchhoff-law-Johnson-noise (KLJN) symmetric-key exchanger.

## 1.2 The KLJN Scheme

The KLJN system [40–99] is a statistical physical scheme based on the Second Law of Thermodynamics. It is a classical (statistical) physical alternative of Quantum Key Distribution (QKD),

---

[3]Auguste Kerckhoffs, not to be confused with Gustav Kirchhoff.

the base of quantum cryptography, which utilizes quantum physical photonic features.

As an illustration of the difficulties and depth of the issues of security, in papers [3–39], important criticisms and attacks are presented about QKD indicating some of the most important aspects of unconditionally secure quantum hardware and their theory. Very recently, National Security Agency (NSA) has made a public statement: "NSA does not recommend the usage of quantum key distribution and quantum cryptography for securing the transmission of data in National Security Systems (NSS)...". While we have reservations about this NSA statement, it inherently brings KLJN up for further considerations. Both the QKD and KLJN schemes count as "Post-Quantum" hardware solutions because they are resilient against any (past and future) computing schemes, including quantum computers, but only the KLJN scheme can be integrated on chips with currently available technology to offer unconditional security of communication within computers and instruments.

Figure 1.2 illustrates the core of the KLJN scheme. The two communicating parties Alice and Bob are connected via a wire. They have identical pairs of resistors $R_A$ and $R_B$. The statistically independent thermal noise voltages $U_{H,A}(t)$, $U_{L,A}(t)$, and $U_{H,B}(t)$, $U_{L,B}(t)$ represent the noise voltages of the resistors $R_H$ and $R_L$ ($R_H > R_L$) of Alice and Bob, respectively, which are generated from random number generators (RNGs) and must have a Gaussian amplitude distribution [59,62].

At the beginning of each bit exchange period (BEP), Alice and Bob randomly choose one of their resistors to connect to the wire. The wire voltage $U_w(t)$ and current $I_w(t)$ are as follows:

$$U_w(t) = I_w(t) R_B + U_B(t), \tag{1.3}$$

$$I_w(t) = \frac{U_A(t) - U_B(t)}{R_A + R_B} \tag{1.4}$$

where $U_A(t)$ and $U_B(t)$ denote the instantaneous noise voltage of the resistor chosen by Alice and Bob, respectively. Alice and Bob (as well as Eve) use the mean-square voltage of the wire to assess the situation. According to the Johnson formula,

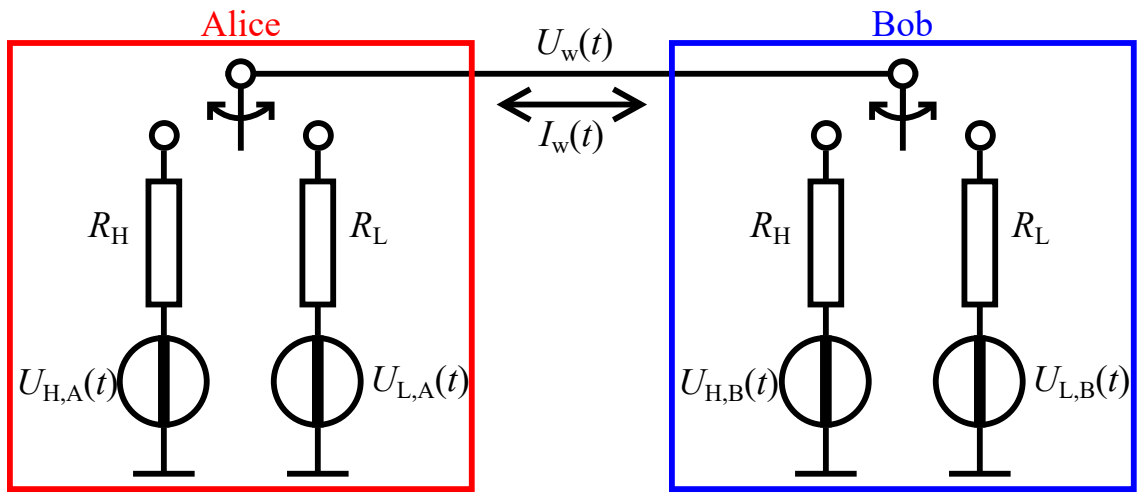

Figure 1.2: The core of the KLJN scheme. The two communicating parties Alice and Bob are connected via a wire. The wire voltage and current are denoted as $U_w(t)$ and $I_w(t)$, respectively. The parties have identical pairs of resistors $R_H$ and $R_L$ ($R_H > R_L$) that are randomly selected and connected to the wire at the beginning of the bit exchange period. The statistically independent thermal noise voltages $U_{H,A}(t)$, $U_{L,A}(t)$, and $U_{H,B}(t)$, $U_{L,B}(t)$ represent the noise voltages of the resistors $R_H$ and $R_L$ of Alice and Bob, respectively.

$$U_{w,\text{eff}}^2 = 4kT_\text{eff} R_p \Delta f_B, \tag{1.5}$$

where $k$ is the Boltzmann constant (1.38 x 10$^{-23}$ J/K), $T_\text{eff}$ is the publicly agreed effective temperature, $R_P$ is the parallel combination of Alice and Bob's chosen resistors, given by

$$R_P = \frac{R_A R_B}{R_A + R_B}, \tag{1.6}$$

and $\Delta f_B$ is the noise bandwidth of the generators.

Four possible resistance situations can be formed by Alice and Bob: HH, LL, LH, and HL. Using the Johnson formula, these correspond to three mean-square voltage levels, as shown in Figure 1.3.

The HH and LL cases represent insecure situations because they form distinct mean-square voltages. The HL and LH cases represent secure bit exchange because Eve cannot distinguish between the corresponding two resistance situations (HL and LH). On the other hand, Alice and

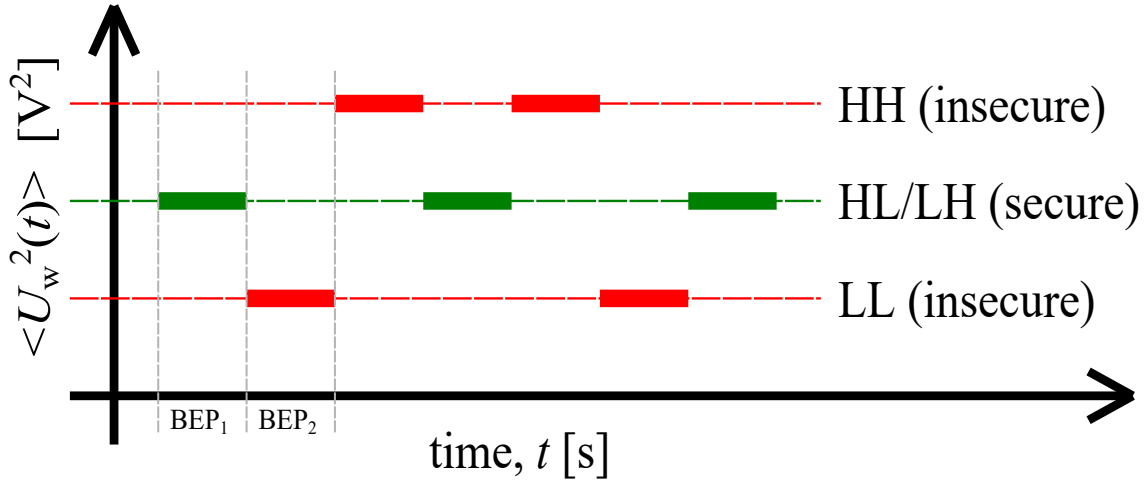

Figure 1.3: The three mean-square voltage levels. The HH and LL cases represent insecure situations because they form distinct mean-square voltages. The HL and LH cases represent secure bit exchange because Eve cannot distinguish between the corresponding two resistance situations (HL and LH). On the other hand, Alice and Bob can determine the resistance at the other end because they know their own connected resistance values.

Bob can determine the resistance at the other end because they know their own connected resistance values.

Several attacks against the KLJN system have been proposed [40–44, 79–99], but no attack has been able to compromise its information-theoretic (unconditional) security because each known attack is either invalid (conceptually incorrect errors in theory and/or experiments) or can be nullified by a corresponding defense scheme such that Eve's information entropy about the key approaches the bit length of the key, while Alice's and Bob's information entropy about the key approaches zero.

The rest of this dissertation is as follows. Chapter 2 introduces four deterministic and four statistical attacks based on the assumption that the random number generators Alice and Bob use to generate their noises are compromised. Chapter 3 introduces an attack based on coincidence events between the wire voltage and current. Chapter 4 introduces an attack based on the nonlinearity of Alice's and Bob's noise generators. Finally, Chapter 5 summarizes and concludes this dissertation.

# 2. RANDOM NUMBER GENERATOR ATTACK AGAINST THE KIRCHHOFF-LAW-JOHNSON-NOISE SECURE KEY EXCHANGER[1,2]

## 2.1 Random Number Generator Attacks

There are two classes of practical random number generators: true (physical) and computational. The nature of computational RNGs is that they collect randomness from various low-entropy input streams and try to generate outputs that are in practice indistinguishable from truly random streams [100–105]. The randomness of an RNG relies on the uncertainty of the random seed, or initialization vector, and a long sequence with uniform distribution. The moment an adversary learns the seed, the outputs are known, and the RNG is compromised.

Various RNG attacks exist against conditionally secure communications [100–105]. Unconditionally secure communications also require true random numbers for perfect security. That is also true for the noises of Alice and Bob, and for the randomness of their switch driving. Yet, it is unclear how Eve can utilize compromised RNGs to attack the KLJN scheme. Here, we demonstrate with simple attack examples that compromised noises lead to information leak.

The rest of this chapter is organized as follows. Section 2.2.1 describes the new deterministic attack protocols, Section 2.2.2 demonstrates their results, and Section 2.2.3 closes this part of the chapter. Then, Section 2.3 describes the new statistical attack protocols, Section 2.3.2 demonstrates their results, and Section 2.3.3 closes this chapter.

---

## 2.2 Deterministic RNG Attack against the KLJN Scheme

### 2.2.1 Attack Methodology

Two theoretical situations are introduced where Eve can use compromised RNGs to crack the KLJN scheme: one where Eve knows the roots of both Alice's and Bob's generators (bilateral parameter knowledge), and another where Eve knows the root of only Bob's generator (unilateral parameter knowledge). A scenario is also introduced where the only statistical evaluation needed is in the rarely-occurring event of a negligible waiting/verification (no-response) time.

*2.2.1.1 Bilateral Parameter Knowledge*

2.2.1.1.1 *Ohm's Law*

Eve knows the roots of both Alice's and Bob's RNGs, thus she knows the instantaneous amplitudes of the noise voltage generators for each of their resistors (see Figure 1.2). With $U_{H,A}(t)$, $U_{L,A}(t)$, $U_{H,B}(t)$, and $U_{L,B}(t)$ known, Eve measures the wire voltage $U_w(t)$ and the wire current $I_w(t)$ to determine $R_B$ (see Equation (1.3)) and determines that the correct waveform for $R_B$ is a flat line at the expected resistor value, while the incorrect waveform is a noise with divergent spikes. This is because the difference between $U_w(t)$ and $U_B(t)$ is directly proportional to $I_w(t)$, with $R_B$ as the scaling factor, and the incorrect noise voltage for $U_B(t)$ would not render a constant value for $R_B$.

A realization of the waveforms for $R_B$ is shown in Figure 2.1 at $R_H = 100$ k$\Omega$, $R_L = 10$ k$\Omega$, $T_{\text{eff}} = 10^{18}$ K, and $\Delta f_B = 500$ Hz. At the present situation, the flat line corresponds to $R_H$, thus Eve determines that Bob has chosen $R_H$. With $U_A(t)$, $U_B(t)$, $R_B$, and $I_w(t)$ known, Eve uses Equation (1.4) to solve for $R_A$.

2.2.1.1.2 *One-Bit Powers*

Eve knows the roots of both Alice's and Bob's RNGs, thus she knows the instantaneous amplitudes of the noise voltage generators for each of their resistors (see Figure 1.2). With $U_{H,A}(t)$, $U_{L,A}(t)$, $U_{H,B}(t)$, and $U_{L,B}(t)$ known, Eve uses Equations (1.3) and (1.4) to come up with the four possible waveforms for $U_w(t)$ and $I_w(t)$ and therefore the four possible waveforms for the instantaneous power flow from Alice to Bob, given by

$$P_{\text{w}}(t) = U_{\text{w}}(t) I_{\text{w}}(t). \tag{2.1}$$

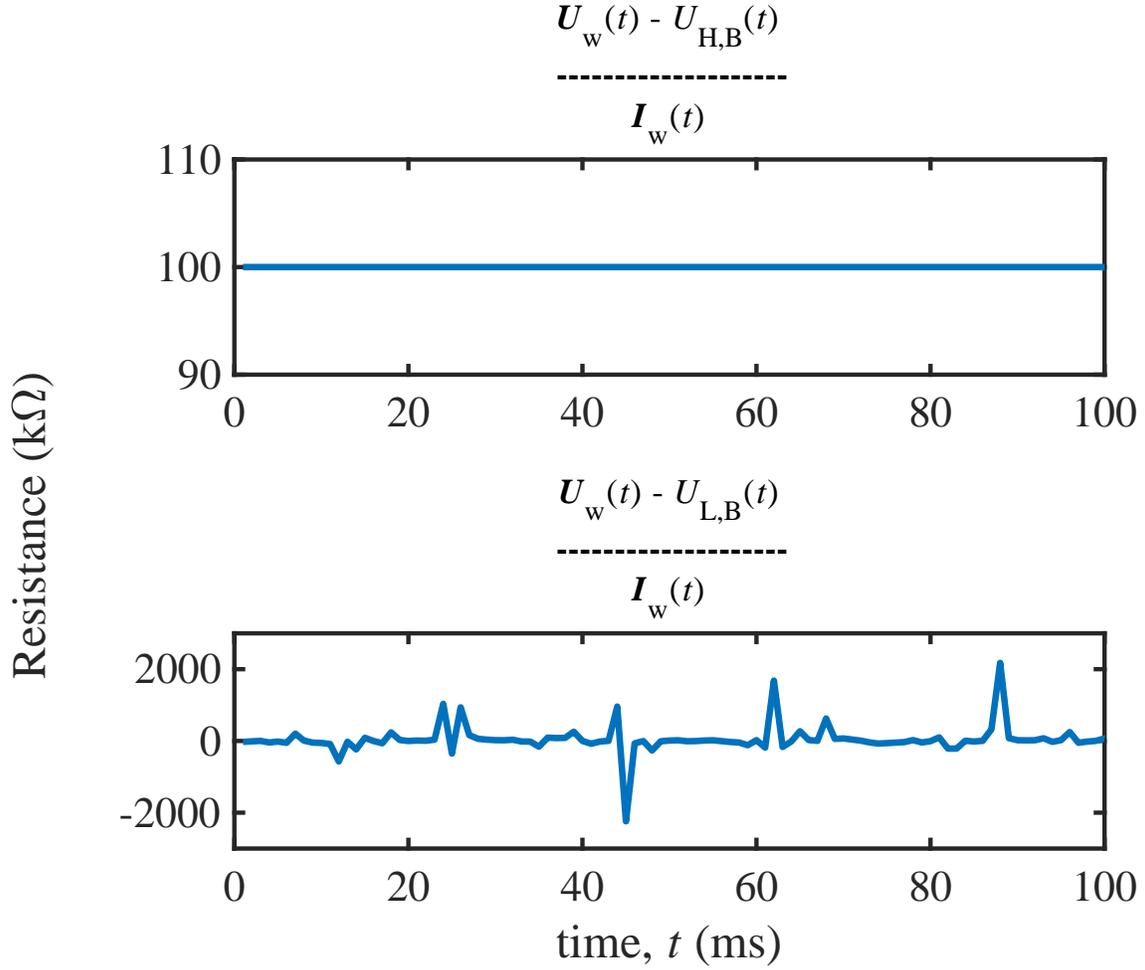

Figure 2.1: A realization of the waveforms for $R_{\text{B}}$ (see Equation (1.3)) at $R_{\text{H}} = 100$ k$\Omega$, $R_{\text{L}} = 10$ k$\Omega$, $T_{\text{eff}} = 10^{18}$ K, and $\Delta f_{\text{B}} = 500$ Hz. The flat line corresponds to $R_{\text{H}}$, thus Eve determines that Bob has chosen $R_{\text{H}}$. The incorrect waveform is a noise with divergent spikes.

Eve measures $U_{\text{w}}(t)$ and $I_{\text{w}}(t)$ and determines the actual $P_{\text{w}}(t)$, which will be identical to one of her four hypothetical waveforms.

Note, the protocol shown above can be simplified to the extreme: Eve's instruments may have just one bit resolution, which is suitable to determine solely the instantaneous direction of the power flow.

### 2.2.1.2  Unilateral Parameter Knowledge

#### 2.2.1.2.1  *Ohm's Law*

Eve knows only the root of Bob's generator RNG, thus she knows merely the noise generator outputs of Bob's resistors, $U_{L,B}(t)$ and $U_{H,B}(t)$. Alice's generator voltages are unknown to her. Eve uses the same protocol as mentioned in the bilateral case (see Section 2.2.1.1) to determine $R_B$, but because Alice's generator voltages are unknown to her, she cannot use Equation (1.4) to solve for $R_A$. Instead, she uses the KLJN protocol as Bob does: use the entire bit exchange period to evaluate the measured mean-squared voltage on the wire. From that value, by using Equation (1.5), she evaluates the parallel resultant $R_P$ of the resistances of Alice and Bob. From $R_P$ and $R_B$, she can calculate $R_A$. This method can still be applied in case Eve is not sure of her result from the bilateral case (see Section 2.2.1.1).

#### 2.2.1.2.2  *Process of Elimination*

Eve knows the root of Alice's RNG. Thus, she knows merely the noise generator outputs of Alice's resistors, $U_{L,A}(t)$ and $U_{H,A}(t)$. Bob's generator voltages are unknown to her. Eve measures the wire voltage $U_w(t)$ and wire current $I_w(t)$. Then, she calculates the hypothetical voltage drops on Alice's possible choice of resistances $R_H$ and $R_L$. With these data, she tests two hypotheses:

**Hypothesis (i):** Alice has chosen $R_L$.

**Hypothesis (ii):** Alice has chosen $R_H$.

With Hypothesis (i), Eve takes the sum

$$U^*_{R_L}(t) = U_w(t) + I_w(t)R_L. \tag{2.2}$$

To test Hypothesis (i) she compares the $U_A(t)$ determined by Equation (2.2) to $U_{L,A}(t)$. If they are identical, then Hypothesis (i) is correct, otherwise Hypothesis (ii) is valid.

Finally, Eve evaluates the measured mean-square voltage on the wire over the bit exchange period. From that value, by using Equation (1.5), she evaluates the parallel resultant $R_P$ of the resistances of Alice and Bob. From $R_\text{P}$ and $R_\text{A}$, she calculates $R_\text{B}$, thus she has cracked the KLJN scheme.

*2.2.1.3 Timing*

In each attack, Eve may or may not know which noise belongs to which resistor, or the noises may provide a voltage-to-current ratio that results in either $R_\text{H}$ or $R_\text{L}$. In these scenarios, there will be a negligible waiting/verification (no-response) time before Eve can differentiate between the noises. The probability of the two independent noises being equal at any given point is

$$p_0 = \frac{1}{2^\Delta}, \tag{2.3}$$

where $\Delta$ is the number of resolution bits.

We can consider the event that the noises run identically for $n$ subsequent steps to be a geometric distribution, with a probability of

$$p = p_0^n, \tag{2.4}$$

where $n = t/\tau$ and $\tau$ is the autocorrelation time, given by the Nyquist Sampling Theorem

$$\tau \approx \frac{1}{2\Delta f_\text{B}}. \tag{2.5}$$

The approximation sign is due to the quantized nature of the noises, as they are constant throughout the bit exchange periods.

While such an identical match between the two noises is taking place, Eve cannot distinguish between the two noises, thus the actual resistance situation (see Figure 1.3) remains secure. However, the exponential decay in Equation (2.4) yields an efficient cracking scheme of the secure key bit value within a short amount of time. In accordance with Equation (2.4), the probability of the two independent seeds running identically is

$$p \approx p_0^{t/\tau} \approx \left(\frac{1}{2^\Delta}\right)^{t/\tau}. \tag{2.6}$$

## 2.2.2 Demonstration

### 2.2.2.1 Johnson Noise Emulation

First, we generated Gaussian band-limited white noise (GBLWN). Precautions were used to avoid aliasing errors, improve Gaussianity, and reduce bias:

(i) At first, using the MATLAB randn() function, $2^{24}$ or 16,777,216 Gaussian random numbers were generated.

(ii) This process was repeated 10 times to generate 10 independent raw noise series, and then an ensemble average was taken out of those 10 series to create one single noise time function with improved Gaussianity and decreased short-term bias.

(iii) Then this time series was converted to the frequency domain by Fast Fourier Transformation (FFT). To get rid of any aliasing error, we opened the real and imaginary portions of the FFT spectrum and doubled their frequency bandwidths by zero padding.

(iv) Finally, we performed an inverse FFT (IFFT) of the zero-padded spectrum to get the time function of the anti-aliased noise.

The real portion of the IFFT result is the band-limited, anti-aliased noise with improved Gaussianity and decreased bias.

The histogram and probability plot for the generated noise are shown in Figures 2.2 and 2.3, respectively, showing that the noise is Gaussian. Figure 2.4 demonstrates that the noise has band-limited, white power density spectrum and it is anti-aliased.

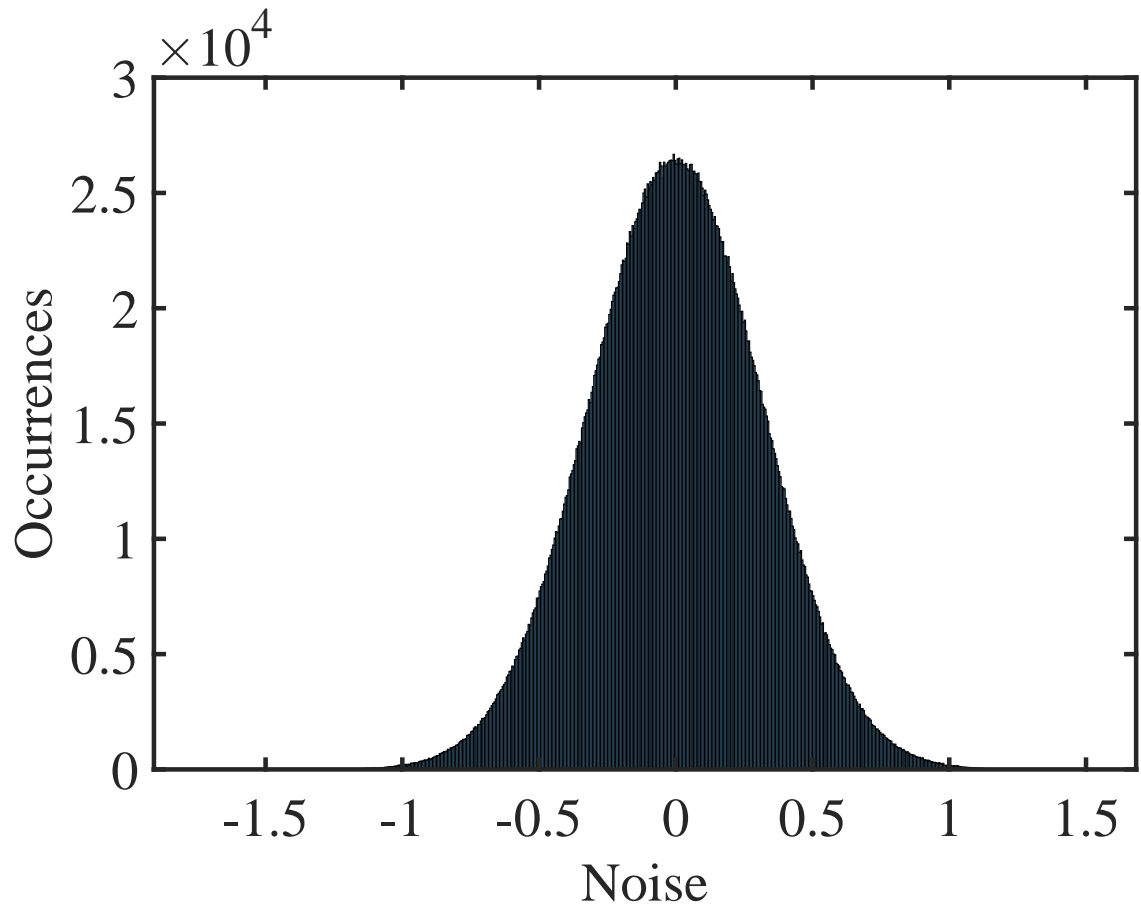

Figure 2.2: Histogram of the noise. A mean of zero and a standard deviation of 1 indicates a standard normal distribution.

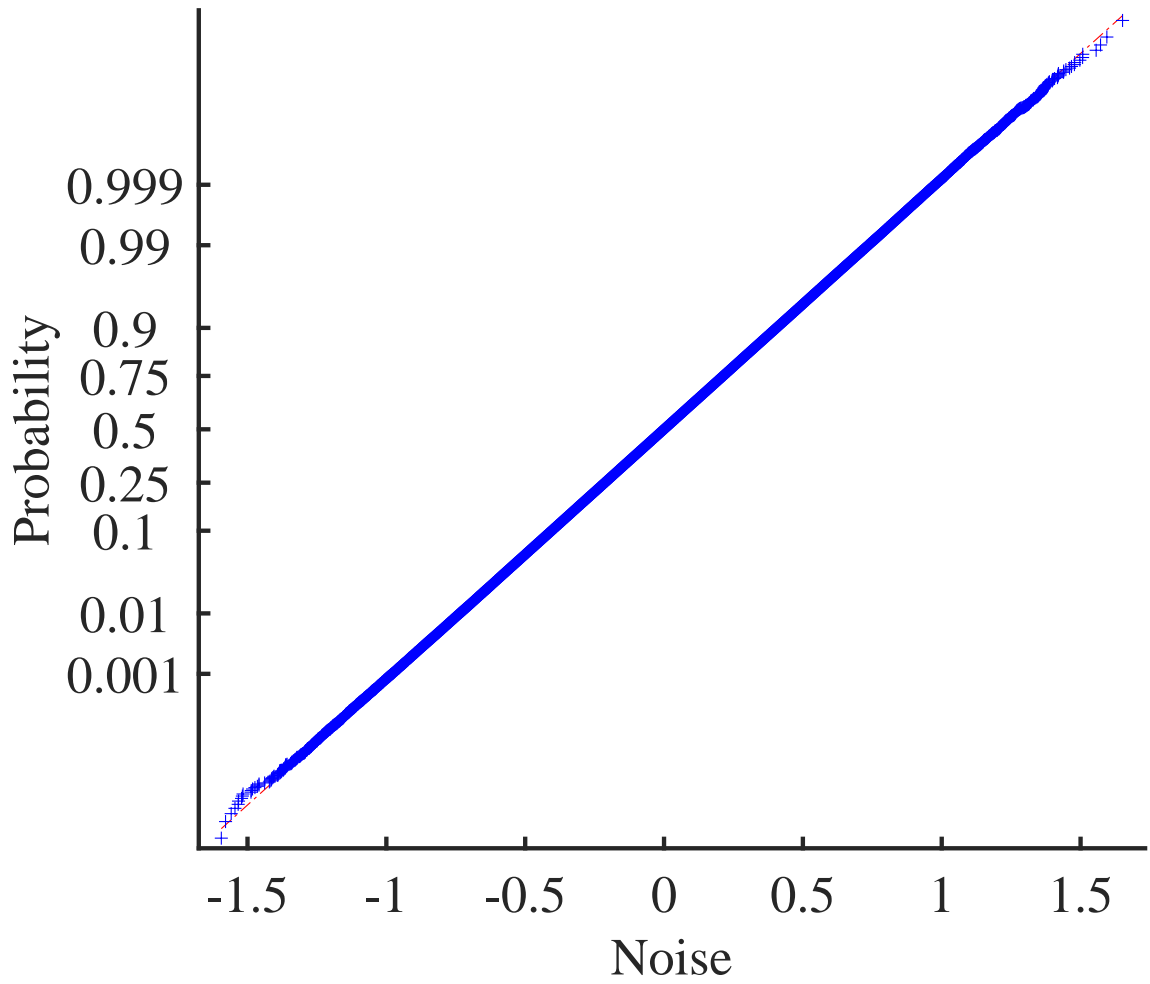

Figure 2.3: Normal-probability plot of the noise. A straight line indicates a pure Gaussian distribution.

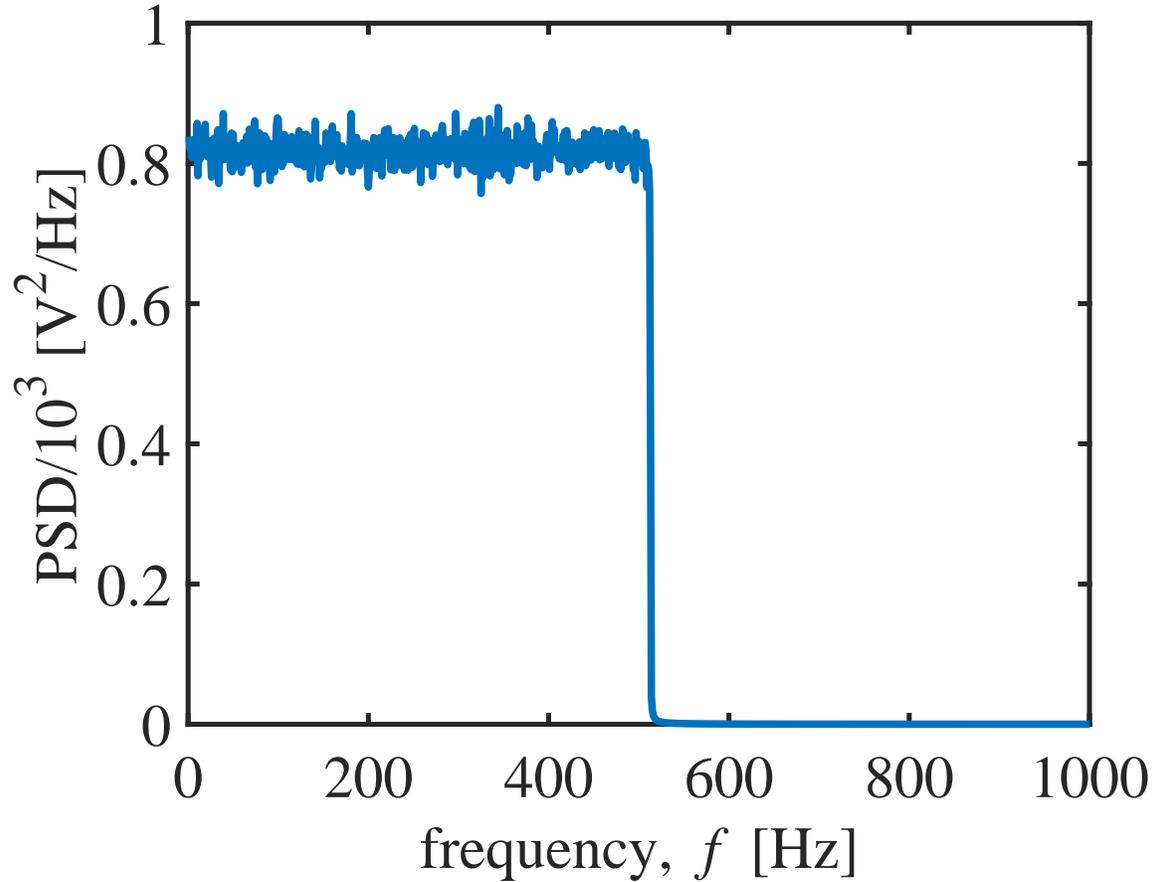

Figure 2.4: Power spectral density of the noise. The bandwidth of the noise is 500 Hz, see Equation (1.5).

The final step was to scale this normalized Gaussian nosie (see Equation (1.5)) to the required level of Johnson noise for the given resistance, temperature, and bandwidth values ($R_\text{L}$, $R_\text{H}$, $T_\text{eff}$, and $\Delta f_\text{B}$, respectively). We chose $R_\text{L}$=10 k$\Omega$, $R_\text{H}$=100 k$\Omega$, $T_\text{eff}$=$10^{18}$ K, and $\Delta f_\text{B}$=500 Hz. A realization of the Johnson noise of $U_{\text{H,A}}(t)$, $U_{\text{L,A}}(t)$, $U_{\text{H,B}}(t)$, and $U_{\text{L,B}}(t)$, where $U_{\text{H,A}}(t)$ is the noise voltage of Alice's $R_\text{H}$, $U_{\text{L,A}}(t)$ is the noise voltage of Alice's $R_\text{L}$, $U_{\text{H,B}}(t)$ is the noise voltage of Bob's $R_\text{H}$, and $U_{\text{L,B}}(t)$ is the noise voltage of Bob's $R_\text{L}$ (see Figure 1.2), over 100 milliseconds is displayed in Figure 2.5. The histogram, probability plot, and power spectral density for those noise voltages are shown in Figures 2.6, 2.7, and 2.8, respectively.

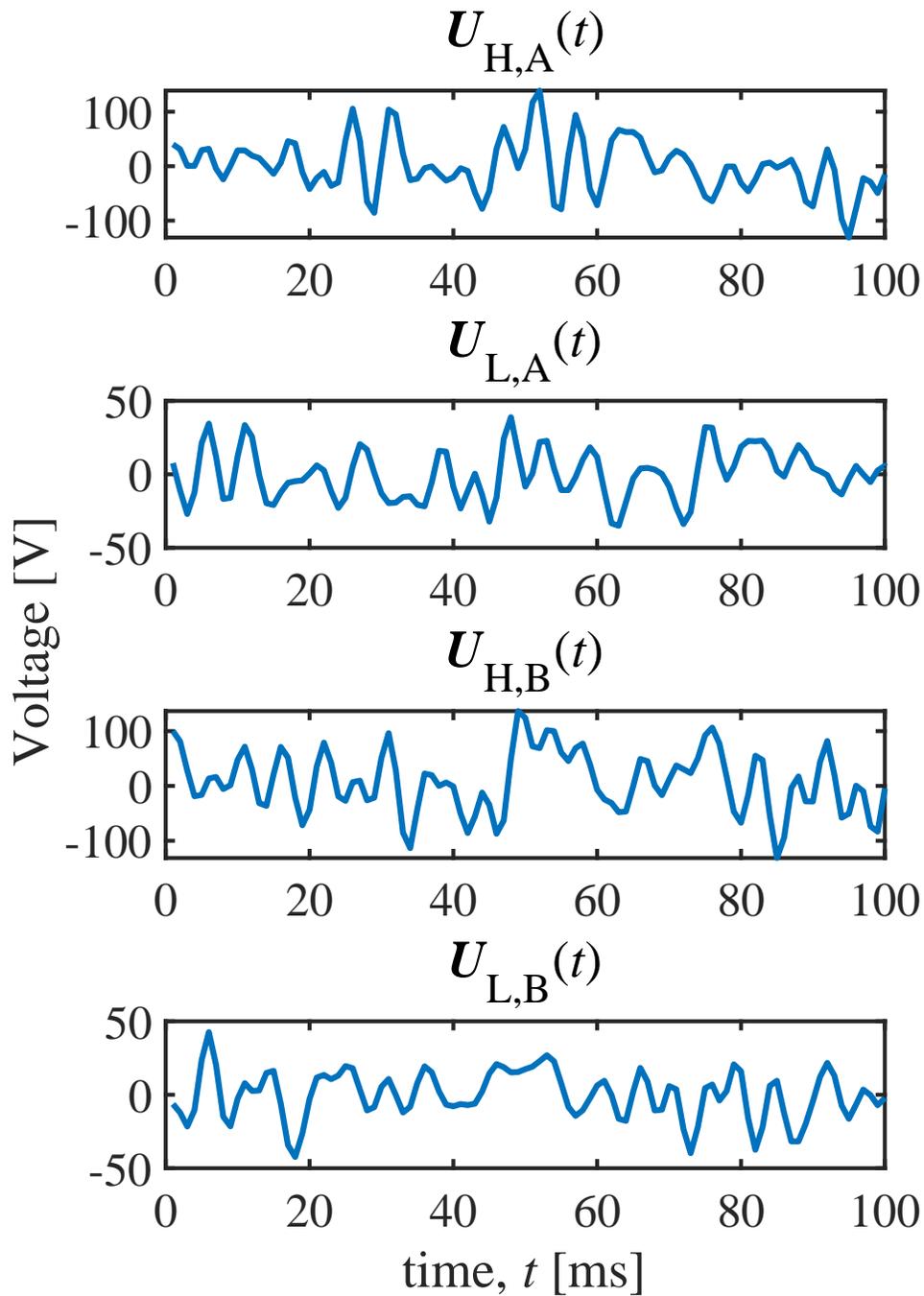

Figure 2.5: A realization of $U_{H,A}(t)$, $U_{L,A}(t)$, $U_{H,B}(t)$, and $U_{L,B}(t)$ (see Figure 1.2) displayed over 100 milliseconds. $U_{H,A}(t)$ is the noise voltage of Alice's $R_H$, $U_{L,A}(t)$ is the noise voltage of Alice's $R_L$, $U_{H,B}(t)$ is the noise voltage of Bob's $R_H$, and $U_{L,B}(t)$ is the noise voltage of Bob's $R_L$. Each time step is one millisecond.

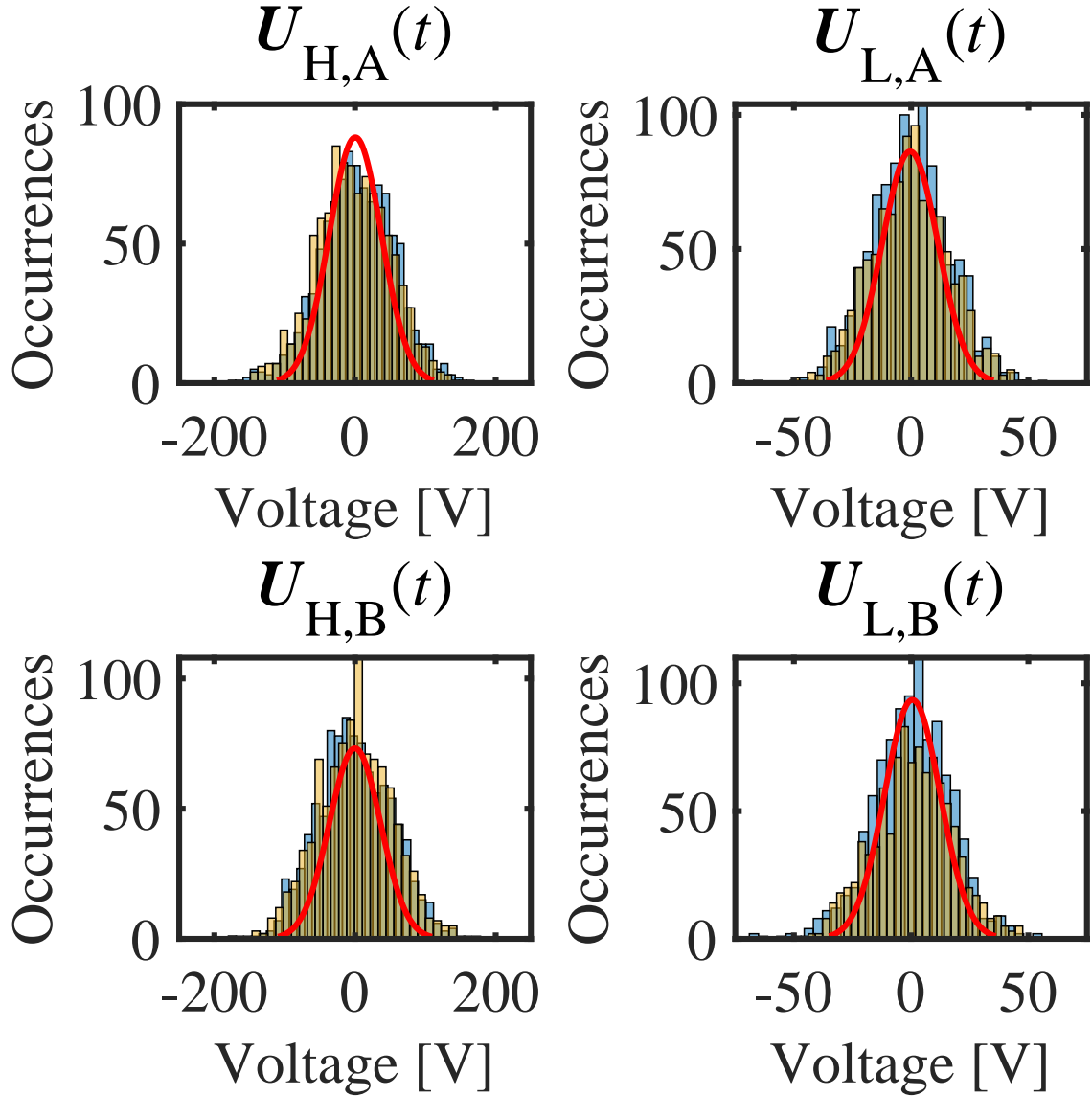

Figure 2.6: Histogram of the Johnson noise of $U_{H,A}(t)$, $U_{L,A}(t)$, $U_{H,B}(t)$, and $U_{L,B}(t)$, where $U_{H,A}(t)$ is the noise voltage of Alice's $R_H$, $U_{L,A}(t)$ is the noise voltage of Alice's $R_L$, $U_{H,B}(t)$ is the noise voltage of Bob's $R_H$, and $U_{L,B}(t)$ is the noise voltage of Bob's $R_L$ (see Figure 1.2). The blue histograms represent the occurrences of the noise voltages over time, and the yellow histograms represent the occurrences of the noise voltages sampled 1000 times at a specific time point. The green area represents the overlap between the blue and yellow histograms. The red line represents the best-fitting normal probability density function (PDF) for both histograms, serving to demonstrate that the noise voltages in all four cases (HH, LL, HL, and LH) are Gaussian and ergodic.

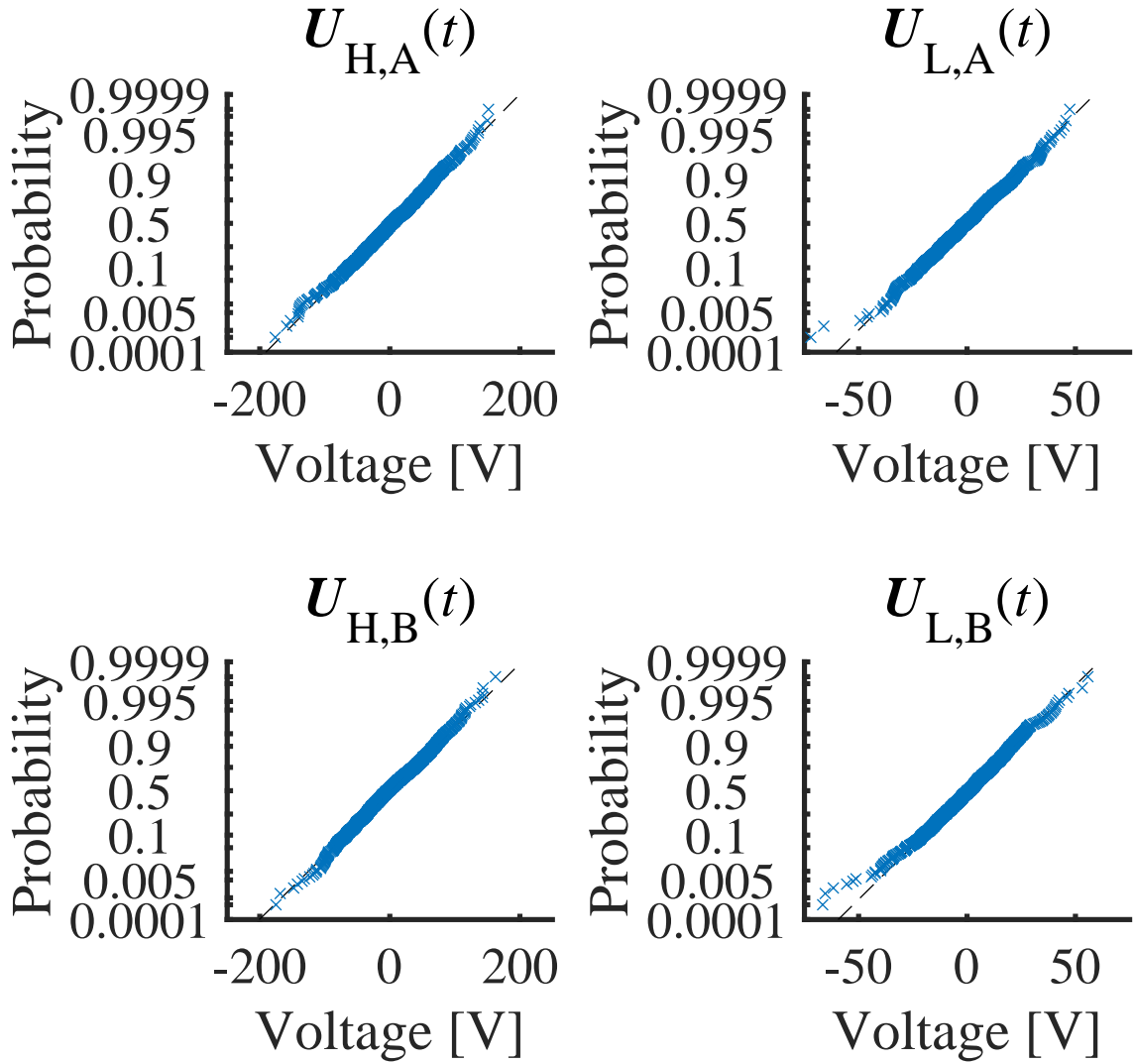

Figure 2.7: Normal-probability plot of the Johnson noise of $U_{H,A}(t)$, $U_{L,A}(t)$, $U_{H,B}(t)$, and $U_{L,B}(t)$. A straight line indicates a pure Gaussian distribution.

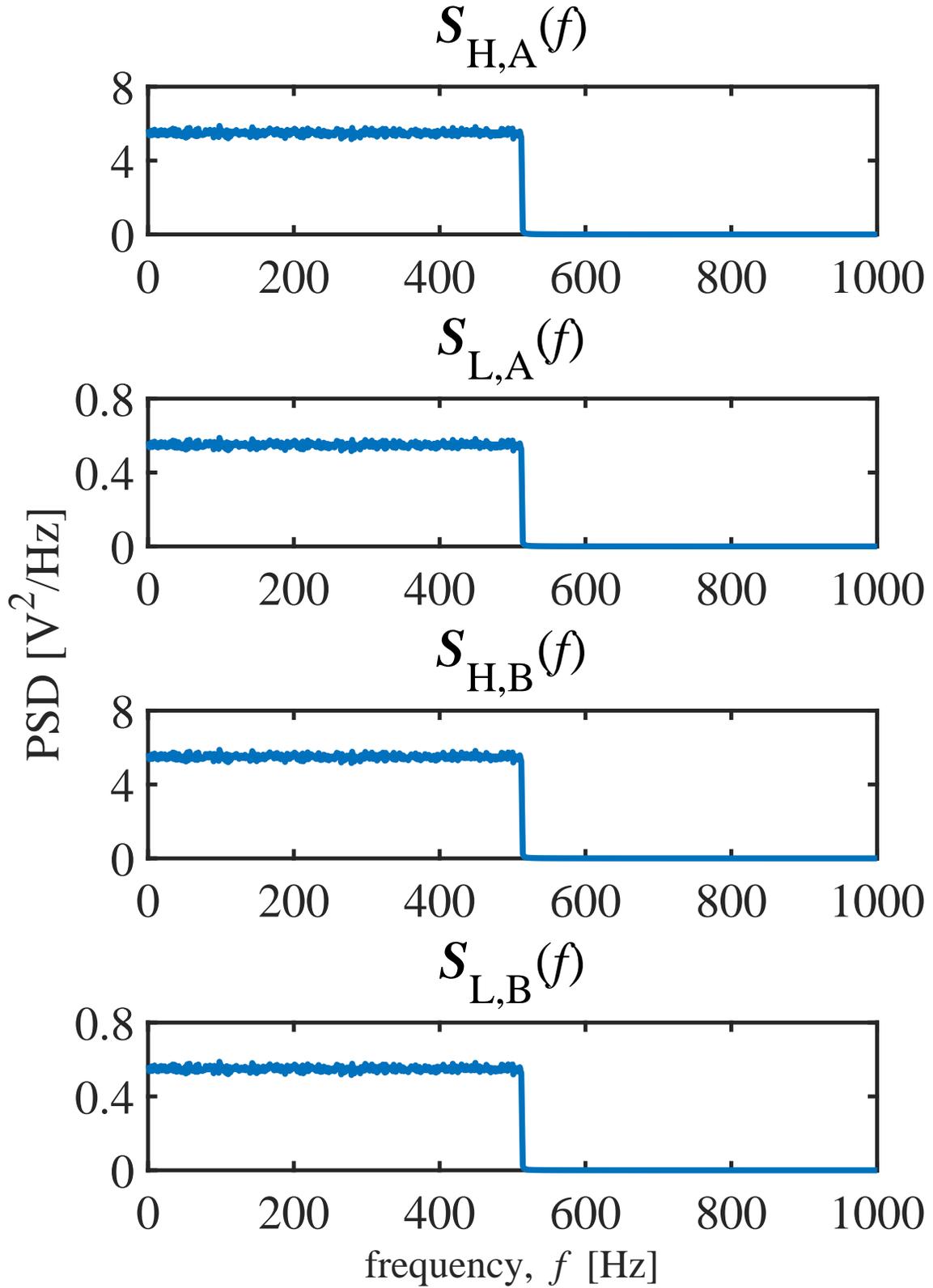

Figure 2.8: Power spectral density of the Johnson noise of $U_{H,A}(t)$, $U_{L,A}(t)$, $U_{H,B}(t)$, and $U_{L,B}(t)$. The bandwidth of the noise is 500 Hz, see Equation (1.5).

From the Nyquist Sampling Theorem,

$$\tau = \frac{1}{2\Delta f_{\text{B}}} \tag{2.7}$$

where $\tau$ represents the time step, a $\Delta f_{\text{B}}$ of 500 Hz renders a time step of $10^{-3}$ seconds. Because this is ideal band-limited white noise, the Nyquist samples are statistically independent.

*2.2.2.2 Setup of Eve's Noises*

To generate Eve's noises that are correlated with the noises in the KLJN system, independent "thermal" noises (see Section 2.2.2.1) are added to the noises of Alice and Bob, and the resulting noises are normalized to have the same effective value as the original noises.

For example, the new correlated noise of Eve corresponding to Alice's thermal noise $U_{\text{L,A}}(t)$ is

$$U_{\text{EL,A}}(t) = U_{\text{L,A}}(t) + MU^*_{\text{EL,A}}(t), \tag{2.8}$$

where $U_{\text{EL,A}}(t)$ is the new correlated noise, $U^*_{\text{EL,A}}(t)$ is an independently generated "thermal" noise for resistance value $R_{\text{L}}$, and the $M$ coefficient controls the cross-correlation between Eve's and Alice's noises, $CCC_{\text{LA}}$ (see Section 2.3.1.1.2). The direct relationship between $M$ and $CCC_{\text{LA}}$ is demonstrated in Tables 2.2 and 2.4, respectively located in Sections 2.3.2.1.2 and 2.3.2.2.2.

Note, $U_{\text{EL,A}}(t)$ and $U^*_{\text{EL,A}}(t)$ have the same effective value. Accordingly,

$$U^*_{\text{EL,A}}(t) = x(t)\sqrt{4kT_{\text{eff}}R_{\text{L}}\Delta f_{\text{B}}}, \tag{2.9}$$

where $x(t)$ represents a new noise with 1 Volt effective value.

However, the effective value of $U_{\text{EL,A}}(t)$ does not satisfy the Johnson formula anymore, so in order to call it a thermal noise, it must be scaled to the proper level [41]:

$$U_{\text{L,A}}(t) = \frac{U_{\text{EL,A}}(t)}{\sqrt{\langle [U_{\text{EL,A}}(t)]^2 \rangle}}\sqrt{4kT_{\text{eff}}R_{\text{L}}\Delta f_{\text{B}}}. \tag{2.10}$$

In this way, the same $M$ used for the different resistance choices of Alice and Bob results in the same cross-correlation coefficient between Eve's noises and the corresponding noises of the communicating parties. Thus, four new sets of independent additive noises have been generated using the same method as in Section 2.2.2.1.

A realization of Eve's noise voltages in comparison to Alice's and Bob's noise voltages over 100 milliseconds is displayed in Figure 2.9. $U_{H,A}(t)$ is the noise voltage of Alice's $R_H$, $U_{L,A}(t)$ is the noise voltage of Alice's $R_L$, $U_{H,B}(t)$ is the noise voltage of Bob's $R_H$, and $U_{L,B}(t)$ is the noise voltage of Bob's $R_L$ (see Figure 1.2). Each time step is one millisecond.

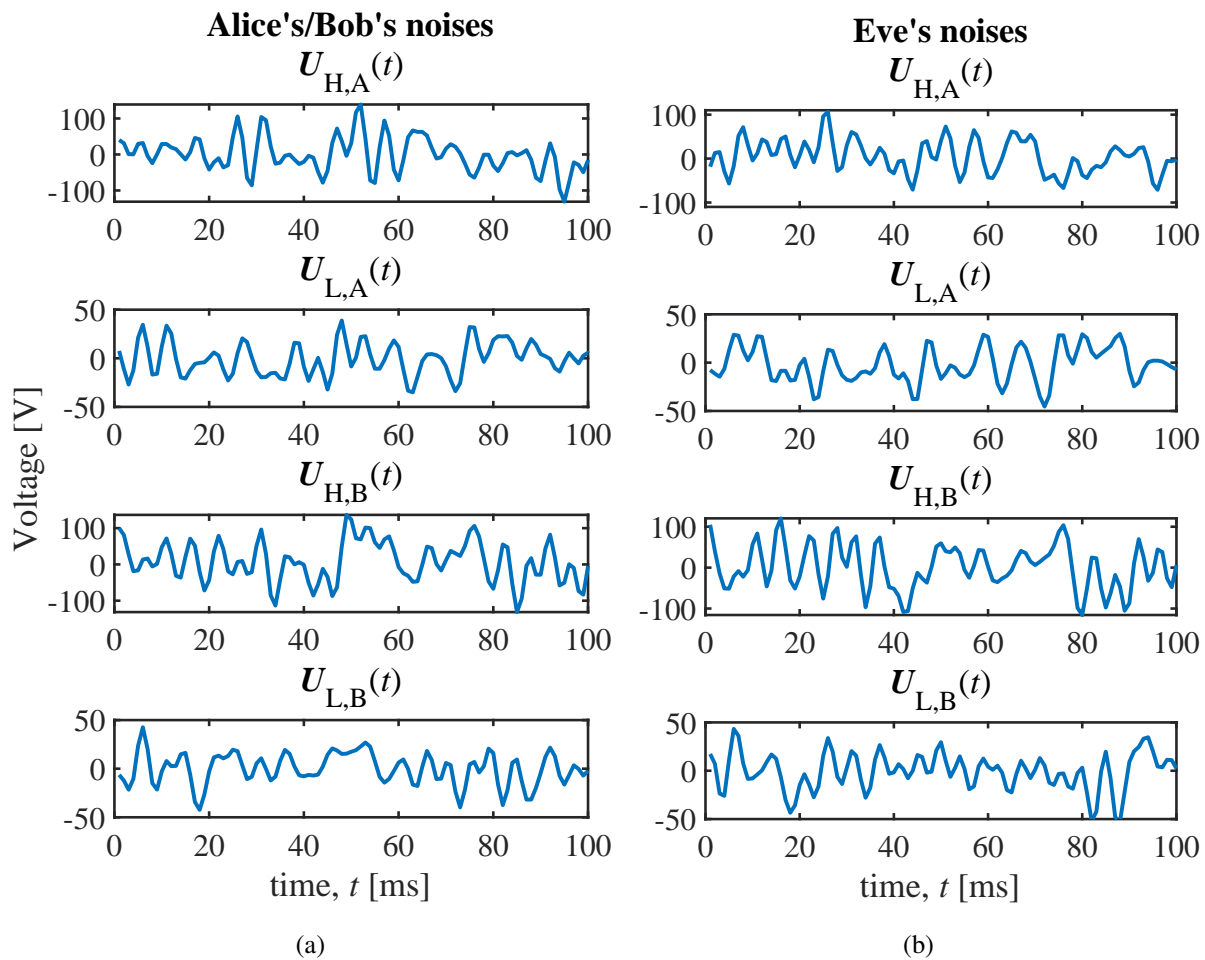

Figure 2.9: A realization of $U_{H,A}(t)$, $U_{L,A}(t)$, $U_{H,B}(t)$, and $U_{L,B}(t)$ (see Figure 1.2) for Alice and Bob (a), and for Eve (b), at $M = 1$, displayed over 100 milliseconds.

For the deterministic attack demonstration (see Section 2.2.2), we keep $M$ at 0, while in the statistical attack (see Section 2.3.2), we vary $M$.

*2.2.2.3  Attack Demonstration when Eve Knows Both Noises*

At the bilateral parameter knowledge (see Section 2.2.1.1), Figure 2.10 shows the hypothetical power flow waveforms generated by Eve (top four graphs). The single-bit plot of her measurement of the actual power flow $P_\text{w}(t)$ is shown by the bottom graph. If the power is delivered from Alice to Bob, the single-bit value is +1, while in the opposite case the result is -1. The orange dashed lines in the upper plots indicate the power flow in the 1-bit resolution limit. At the present situation, Eve's measured single-bit data are identical to the dashed lines in the LH scenario only, thus Eve decides that the LH is the secure resistance situation.

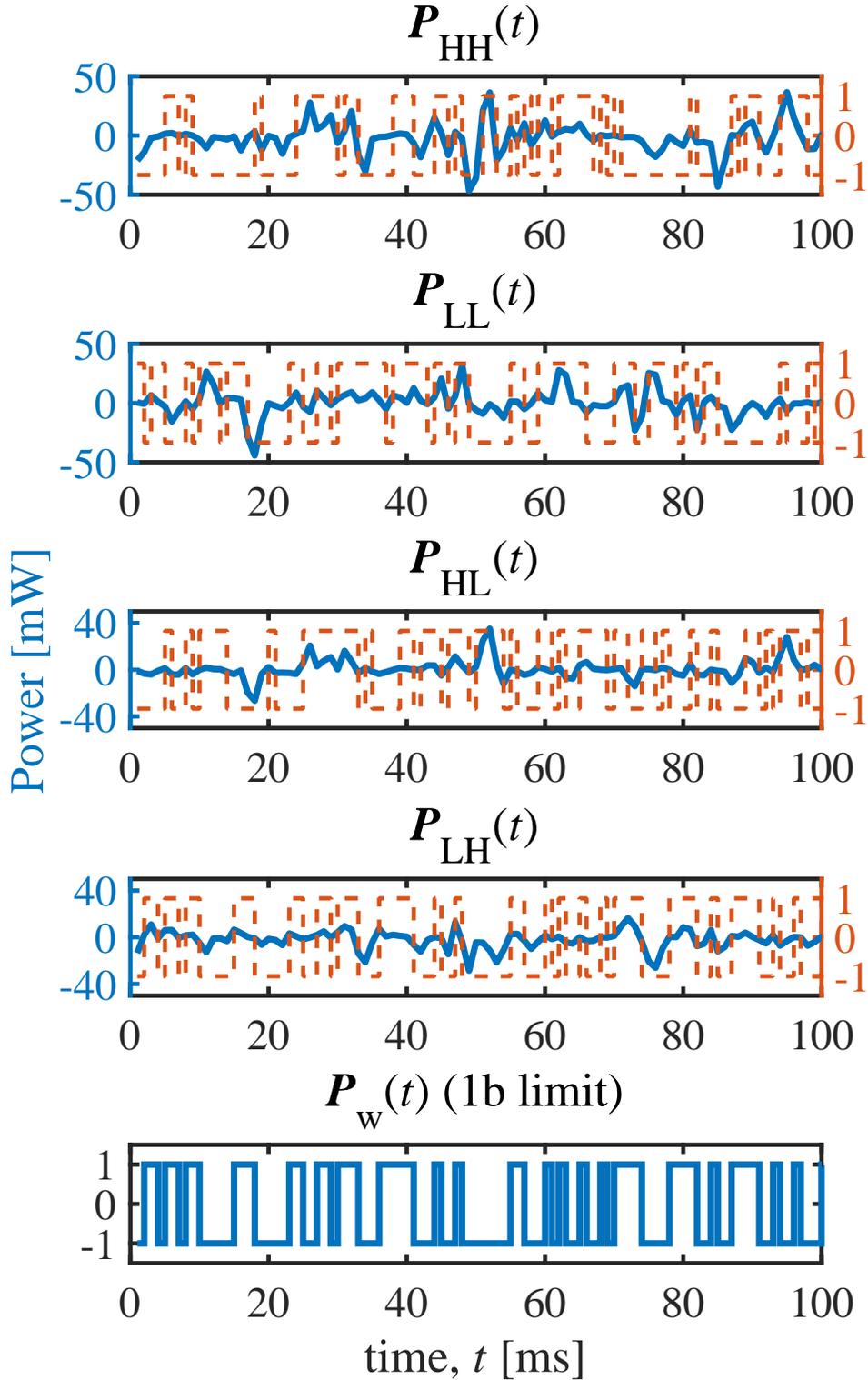

Figure 2.10: The hypothetical and their 1-bit limit waveforms for $P_\text{w}(t)$ generated by Eve. $P_\text{HH}(t)$ is the hypothetical power flow in the HH case, $P_\text{LL}(t)$ is the hypothetical power flow in the LL case, $P_\text{HL}(t)$ is the hypothetical power flow in the HL case, and $P_\text{LH}(t)$ is the hypothetical power flow in the LH case. $P_\text{w}(t)$ is the single-bit measurement of the actual power flow.

However, similarly to the case of string verification with noise-based logic [106], two independent random bit sequences, such as two different binary sequences in Figure 2.10, may run identically for $n$ subsequent steps with probability

$$p = \frac{1}{2^n} \qquad (2.11)$$

where the independence of the subsequent samples within a given binary sequence is also essential, and $n = t/\tau$.

While such an identical match between two of the 4 noises is taking place, Eve cannot decide which one of the hypothetical sequences is valid thus the actual resistor situation (see Figure 1.3) remains secure. However, the exponential decay of the probability of the duration of this phenomenon in Equation (2.11) yields an efficient cracking of the secure key bit value within a short time. In accordance with (2.11), the probability of two independent ones of our binary sequences (Figure 2.10) running identically is

$$p(t) \approx \frac{1}{2^{t/\tau}} = \frac{1}{2^{1000t}} \qquad (2.12)$$

The approximation sign is due to the quantized nature of $p(t)$ because it is staying constant during the bit exchange periods.

This simulation was run 1000 times. Figure 2.11a shows the probability that the exchange of the current key bit is still secure. Figure 2.11b shows the probability that Eve has already cracked the bit. The red lines follow the scaling given by Equation (2.12).

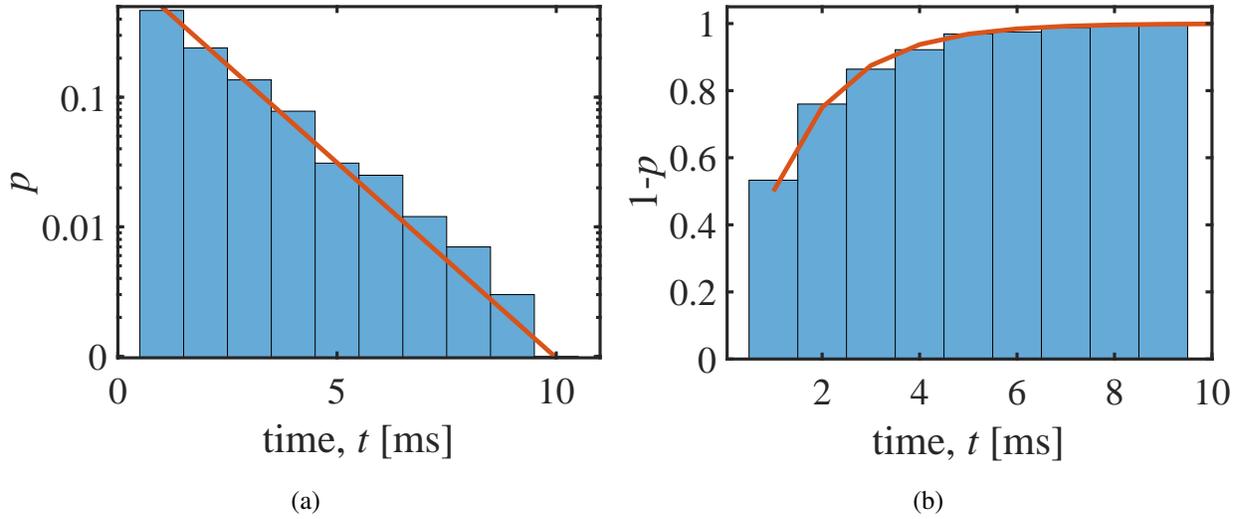

Figure 2.11: Probabilities during the attack with single-bit resolution of Eve's measurements. (a) Probability that the exchange of the current key bit is still secure; (b) and probability that Eve has already cracked it. The histograms represent the simulation results, and the red lines represent the scaling given by Equation (2.12).

*2.2.2.4 Attack Demonstration when Eve Knows Only One of the Sources*

At the unilateral parameter knowledge (see Section 2.2.1.2), Eve knows the root of Alice's RNG, while the root of Bob's RNG is unknown to her. Thus, we suppose she knows $U_A(t)$ (like at the bilateral case), but not $U_B(t)$. A realization of the wire voltage and current, $U_w(t)$ and $I_w(t)$, under the LH condition over 100 milliseconds, is displayed in Figure 2.12. Figure 2.13 shows the hypothetical noise voltages generated by Eve's simulations across $R_L$ and $R_H$ via Ohm's Law.

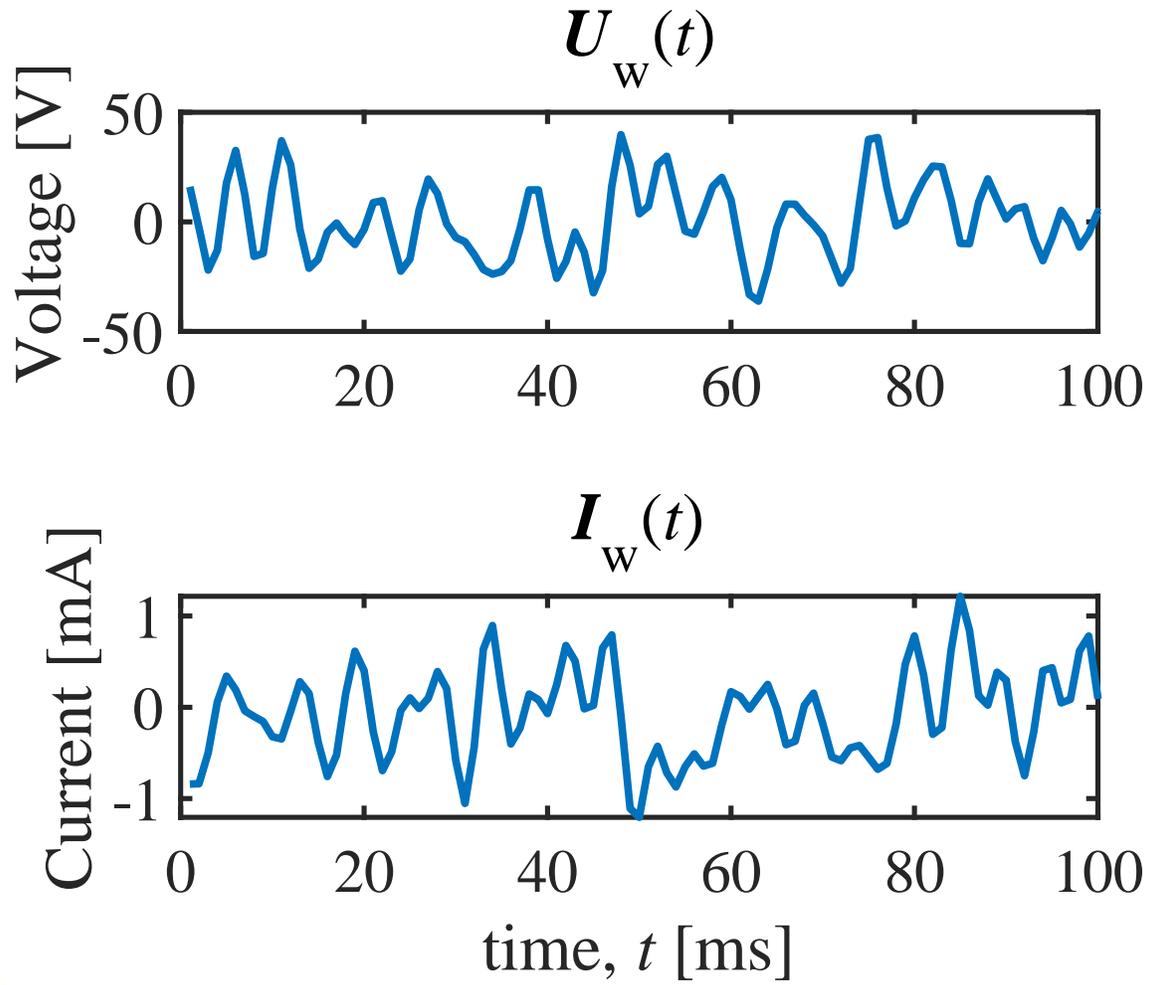

Figure 2.12: A realization of $U_w(t)$ and $I_w(t)$ (see Figure 1.2) displayed over 100 milliseconds. Eve measures and records these data.

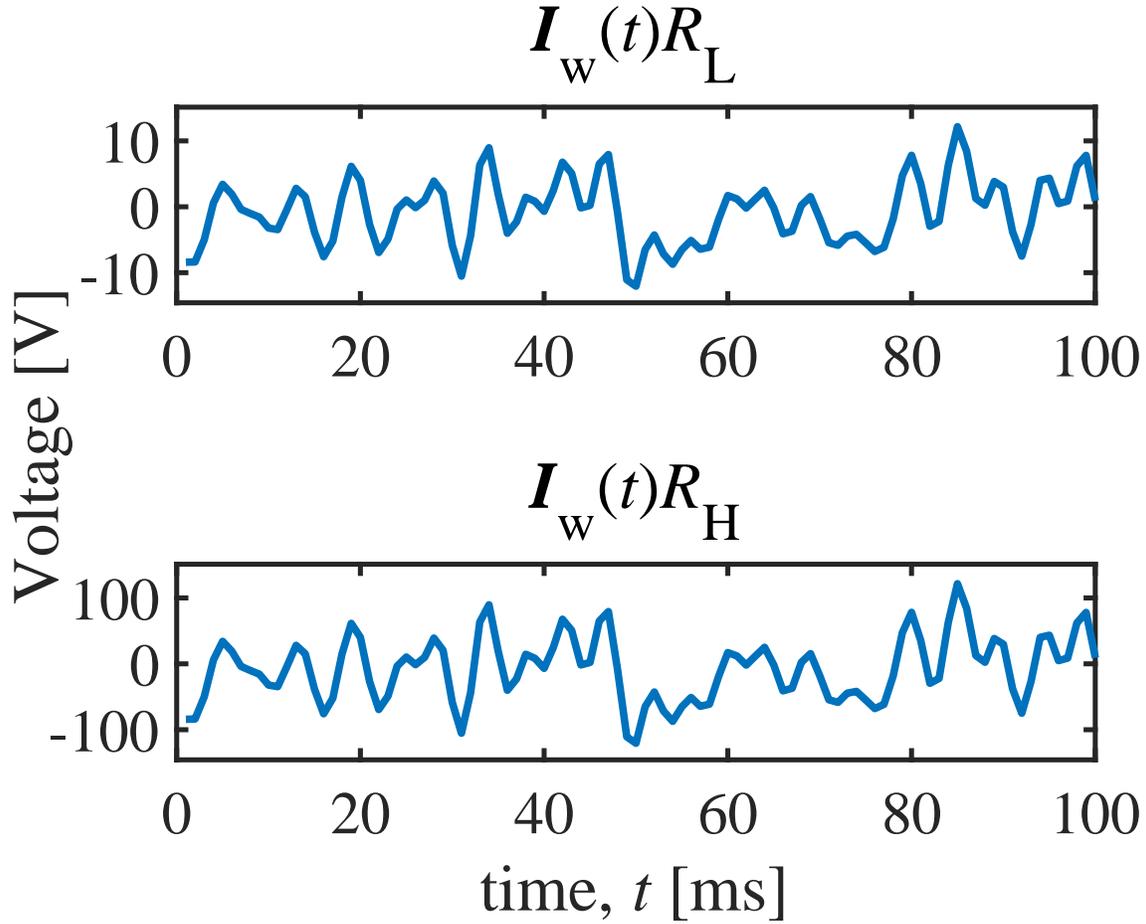

Figure 2.13: Hypothetical noise voltage drops across $R_L$ and $R_H$ by Eve's current measurements and Ohm's law. These results are used in Equation (2.2) to calculate the hypothetical waveforms for $U_{L,A}(t)$ and $U_{H,A}(t)$.

Figure 2.14 shows Eve's hypothetical waveforms for $U_{L,A}(t)$ and $U_{H,A}(t)$, which we denote as $U^*_{R_L}(t)$ and $U^*_{R_H}(t)$ (see Equation (2.2)), in comparison to her known $U_{L,A}(t)$ and $U_{H,A}(t)$. In this case, the waveform for $U^*_{R_L}(t)$ is identical to that of $U_{L,A}(t)$, thus Eve decides that Alice has chosen $R_L$. A correlation plot is also shown in Figure 2.15, between (a) $U^*_{R_L}(t)$ vs. $U_{L,A}(t)$ and $U_{H,A}(t)$, and (b) $U^*_{R_H}(t)$ vs. $U_{L,A}(t)$ and $U_{H,A}(t)$. A one-to-one correlation between $U^*_{R_L}(t)$ and $U_{L,A}(t)$ indicates that Alice has chosen $R_L$.

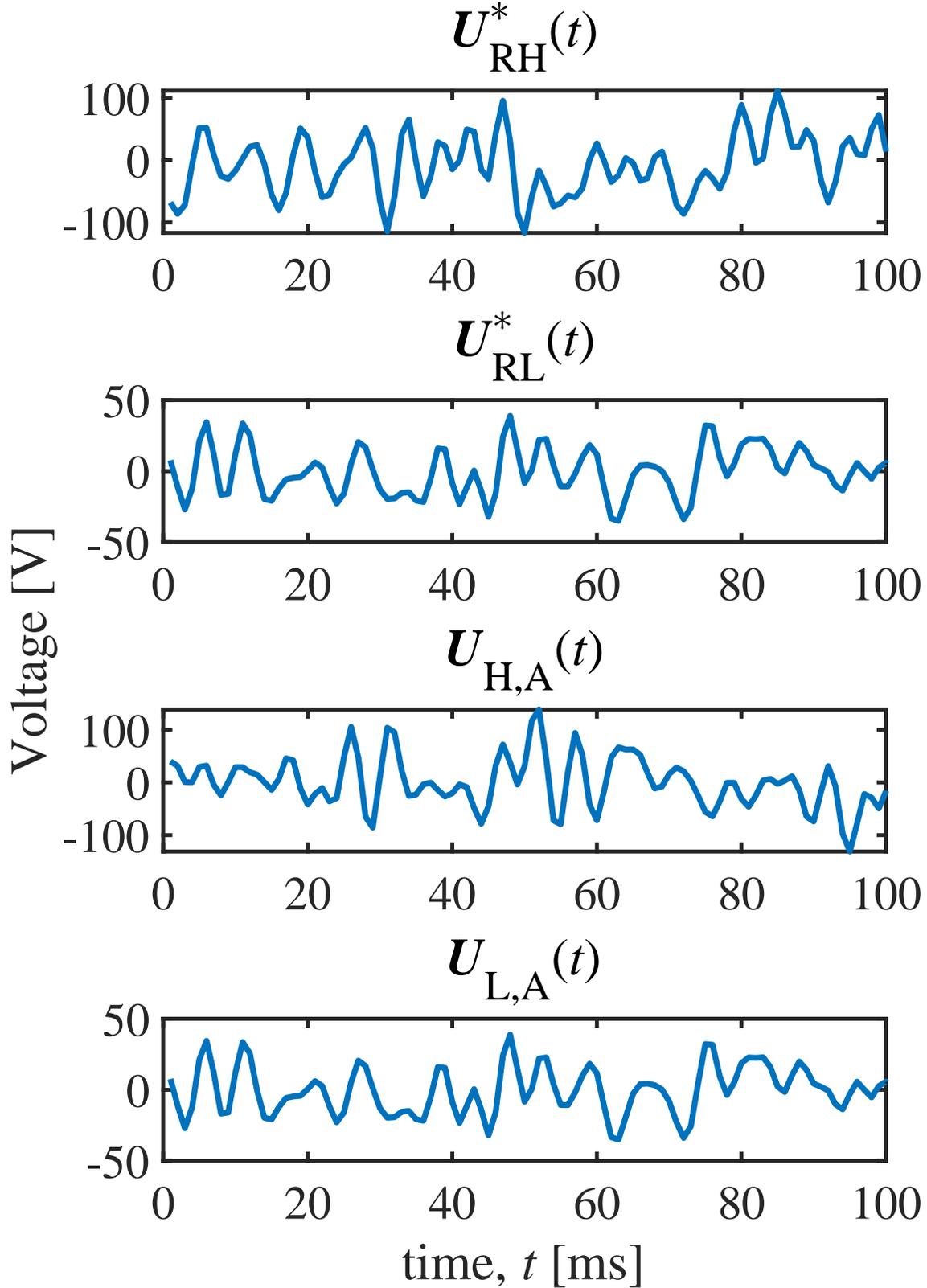

Figure 2.14: Eve's hypothetical waveforms for $U_{L,A}(t)$ and $U_{H,A}(t)$, which we denote as $U^*_{R_L}(t)$ and $U^*_{R_H}(t)$ (see Equation (2.2)), in comparison to her known $U_{L,A}(t)$ and $U_{H,A}(t)$.

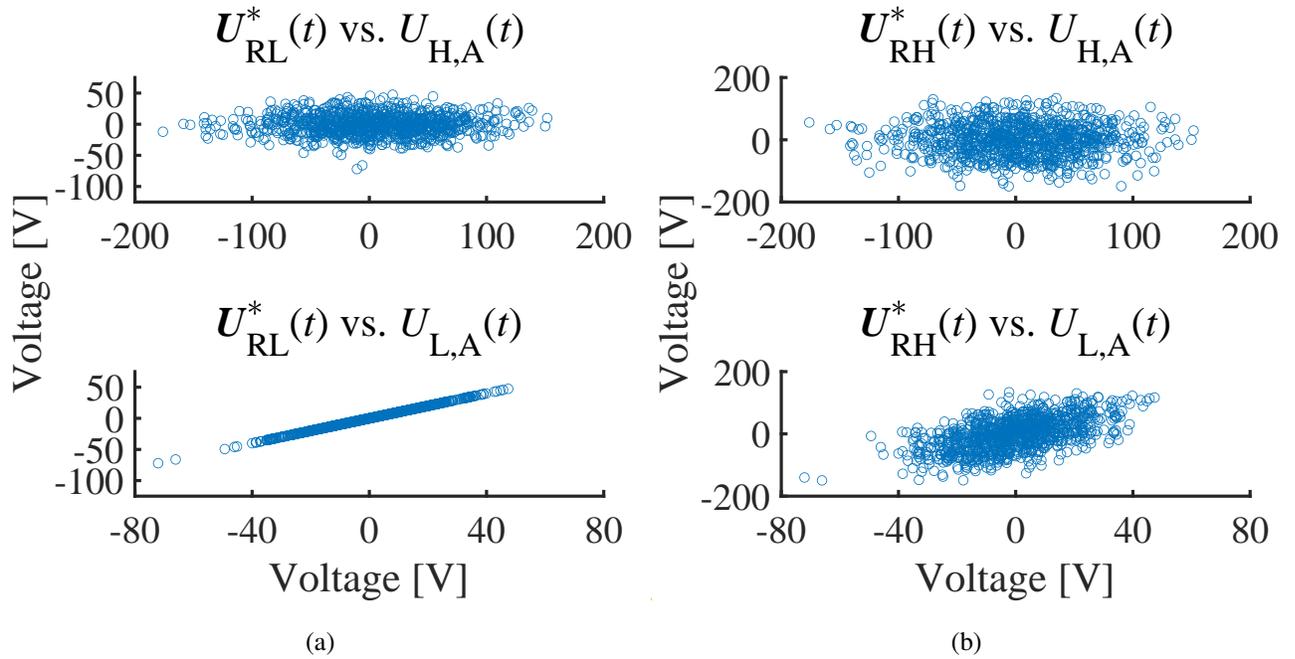

Figure 2.15: Correlation between (a) $U^*_{R_L}(t)$ vs. $U_{L,A}(t)$ and $U_{H,A}(t)$, and (b) $U^*_{R_H}(t)$ vs. $U_{L,A}(t)$ and $U_{H,A}(t)$. A one-to-one correlation between Eve's known $U_A(t)$ and the hypothetical waveform for $U_{L,A}(t)$ indicates that Alice has chosen $R_L$.

Finally, Eve evaluates the measured mean-square voltage on the wire over the bit exchange period, shown in Figure 2.16. From that value, by using Equation (1.5), she evaluates the parallel resultant $R_P$ of the resistances of Alice and Bob. From $R_P$ and $R_A$, she can calculate $R_B$. In this particular case, Bob has chosen $R_H$, thus Eve has cracked the LH situation and the related secure bit value.

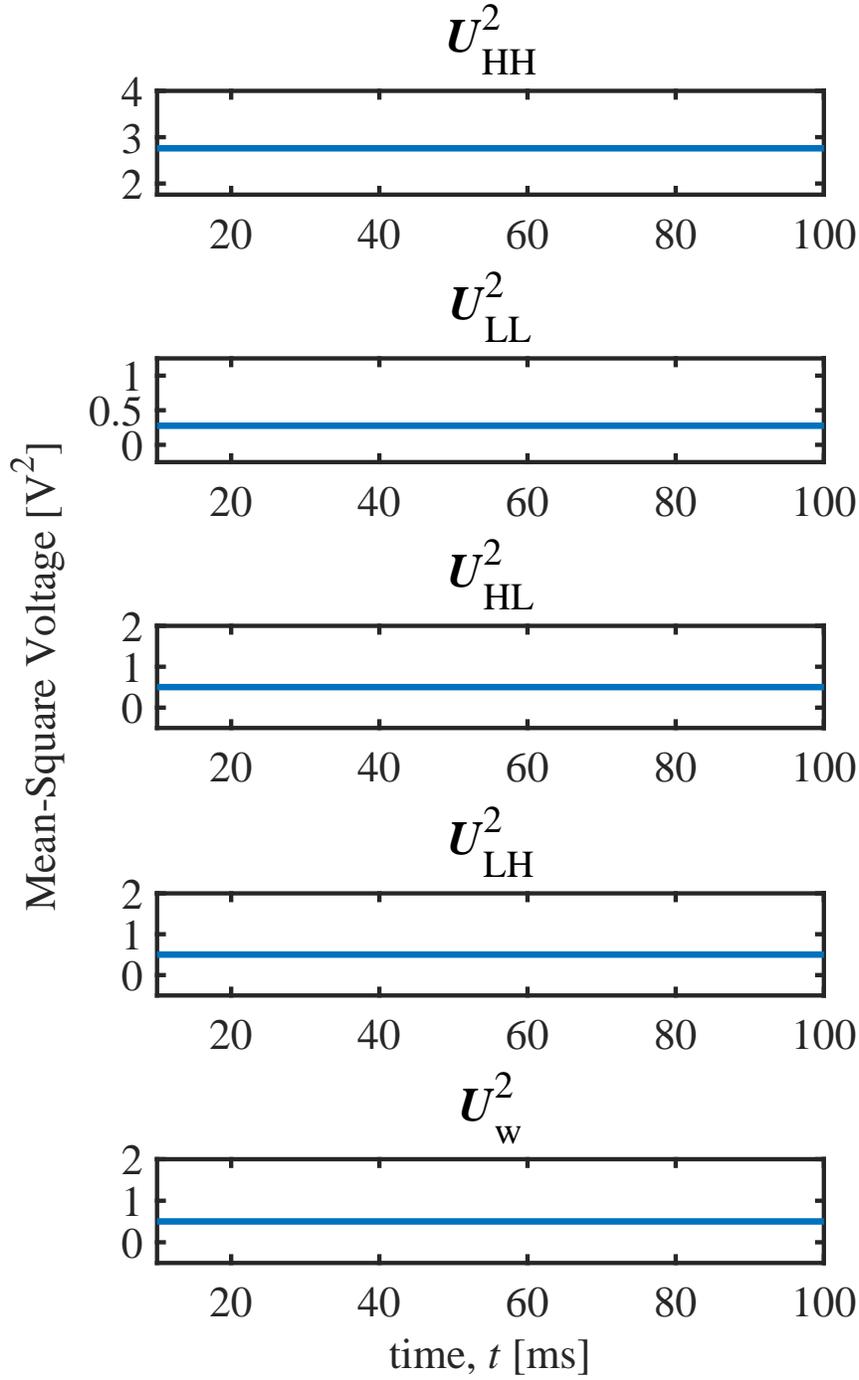

Figure 2.16: The hypothetical and measured mean-square voltage generated by Eve. $U^2_{\text{HH,eff}}$ is the hypothetical mean-square voltage in the HH case, $U^2_{\text{LL,eff}}$ is the hypothetical mean-square voltage in the LL case, $U^2_{\text{HL,eff}}$ is the hypothetical mean-square voltage in the HL case, and $U^2_{\text{LH,eff}}$ is the hypothetical mean-square voltage in the LH case. $U^2_{\text{w,eff}}$ is the measurement of the actual mean-square voltage.

### 2.2.3 Transition

This concludes the deterministic RNG attacks presented in this dissertation. Now, we move onto the statistical RNG attacks.

## 2.3 Statistical RNG Attack against the KLJN Scheme

### 2.3.1 Statistical Attack Protocol

Two situations are introduced where Eve can use compromised RNGs to crack the KLJN scheme: one where Eve has partial knowledge of both Alice's and Bob's generators (bilateral knowledge), and another where Eve has partial knowledge of only Alice's generator (unilateral knowledge).

*2.3.1.1 Bilateral Knowledge*

Eve has statistical knowledge of the amplitudes of the noise voltage generators $U_{\text{H,A}}(t)$, $U_{\text{L,A}}(t)$, $U_{\text{H,B}}(t)$, and $U_{\text{L,B}}(t)$ for each of Alice's and Bob's resistors, see Figure 1.2. These noises are correlated with Alice's and Bob's corresponding noises. Eve then uses Equations (1.3), (1.4), and (2.1) for the cross-correlation attacks shown below.

2.3.1.1.1 *Cross-correlation attack utilizing Alice's/Bob's and Eve's channel voltages, currents and power*

For the four possible resistor combinations, HL, LH, LL and HH, Eve sets up a simulator utilizing her noises (that are correlated with Alice's Bob's corresponding noises), and she records the resulting wire voltages $U_{\text{HH}}(t)$, $U_{\text{LL}}(t)$, $U_{\text{HL}}(t)$, and $U_{\text{LH}}(t)$. Then she evaluates the cross-correlation between the measured wire voltage $U_{\text{w}}(t)$ of Alice and Bob, and the simulated ones. For example, in the HH simulated resistor situation, this is given by

$$CCC_{\text{HH,U}} = \frac{\langle U_{\text{w}}(t) U_{\text{HH}}(t) \rangle}{U_{\text{w}} U_{\text{HH}}}, \qquad (2.13)$$

where $CCC_{\text{HH,U}}$ is the cross-correlation coefficient between her simulated wire voltage $U_{\text{HH}}(t)$ in the HH case and the *measured* channel voltage $U_{\text{w}}(t)$.

Finally, Eve guesses that the actual resistor situation is the one with the highest cross-correlation between the voltage on the wire and in Eve's probing system.

Note, this protocol of cross-correlating the simulated and real wire data can also be repeated for the channel current $I_w(t)$ and channel power $P_w(t)$.

For the channel current protocol, she measures $I_w(t)$ and evaluates the cross-correlation between the measured $I_w(t)$ and her four simulated wire currents $I_{HH}(t)$, $I_{LL}(t)$, $I_{HL}(t)$, and $I_{LH}(t)$. In the HH example, this is given by

$$CCC_{HH,I} = \frac{\langle I_w(t)I_{HH}(t)\rangle}{I_w I_{HH}}, \quad (2.14)$$

where $CCC_{HH,I}$ is the cross-correlation coefficient between her simulated wire voltage $I_{HH}(t)$ in the HH case and the *measured* channel current $I_w(t)$.

Finally, similarly to the above voltage correlation attack, Eve guesses that the bit situation is the one with the highest cross-correlation between the current in the wire and in Eve's probing system.

For the channel power protocol, Eve has the measured $U_w(t)$ and $I_w(t)$ at her disposal. She uses her noises to simulate the four possible waveforms, $P_{HH}(t)$, $P_{LL}(t)$, $P_{HL}(t)$, and $P_{LH}(t)$, for the instantaneous power flow from Alice to Bob, and cross-correlates the with the actual measured $P_w(t)$. In the HH example:

$$CCC_{HH,P} = \frac{\langle P_w(t)P_{HH}(t)\rangle}{P_w P_{HH}}, \quad (2.15)$$

where $P_{HH}(t)$ is the simulated channel power in the HH case, and the measured power is given by Equation (2.1).

Finally, Eve guesses that the bit situation is the one with the highest cross-correlation between the power flows in the wire and in Eve's probing system.

2.3.1.1.2 *Cross-correlation attack directly utilizing Alice's/Bob's and Eve's voltage sources*

There is an alternative way for correlation attacks. Eve measures the wire voltage $U_w(t)$ and wire current $I_{\text{w}}(t)$ [41, 42]. Then, from $I_{\text{w}}(t)$, she uses Ohm's law to calculate the hypothetical voltage drops on Alice's/Bob's possible choices of resistances $R_{\text{H}}$ and $R_{\text{L}}$. From that voltage drop, by using Kirchhoff's loop law, Eve calculates Alice's/Bob's hypothetical noise voltage amplitudes. With these data, she tests four hypotheses:

**Hypothesis (i):** Alice has chosen $R_{\text{L}}$

**Hypothesis (ii):** Alice has chosen $R_{\text{H}}$

**Hypothesis (iii):** Bob has chosen $R_{\text{L}}$

**Hypothesis (iv):** Bob has chosen $R_{\text{H}}$

With Hypothesis (i), to utilize Kirchhoff's loop law, using the same current direction (Alice $\rightarrow$ Bob) as with the power calculation in Equation (2.1), Eve takes the sum

$$U^*_{\text{L,A}}(t) = U_{\text{w}}(t) + I_{\text{w}}(t) R_{\text{L}} \qquad (2.16)$$

to calculate the hypothetical value $U^*_{\text{L,A}}(t)$ of Alice's noise $U_{\text{L,A}}(t)$.

To test Hypothesis (i) she finds the cross-correlation between $U_{\text{L,A}}(t)$ (see Figure 1.2) and the $U^*_{\text{L,A}}(t)$ determined by (Equation (2.16)), given by

$$CCC_{\text{LA}} = \frac{\langle U^*_{\text{L,A}}(t) U_{\text{L,A}}(t) \rangle}{U^*_{\text{L,A}} U_{\text{L,A}}}. \qquad (2.17)$$

Then, with the same equation, she finds the cross-correlation coefficient between $U^*_{\text{L,A}}(t)$ and $U_{\text{H,A}}(t)$. If the former result yields the higher cross-correlation, then Eve has determined that Hypothesis (i) is correct; otherwise, Hypothesis (ii) is valid.

Similarly, at Bob's side, with Hypothesis (iii), Eve takes the difference

$$U^*_{\text{L,B}}(t) = U_{\text{w}}(t) - I_{\text{w}}(t) R_{\text{L}} \qquad (2.18)$$

to calculate the hypothetical value $U^*_{\text{L,B}}(t)$ of Bob's noise $U_{\text{L,B}}(t)$.

To test Hypothesis (iii) she finds the cross-correlation between $U_{\text{L,B}}(t)$ (see Figure 1.2) and the $U_{\text{L,B}}^*(t)$ determined by Equation (2.18), given by

$$CCC_{\text{LB}} = \frac{\langle U_{\text{L,B}}^*(t) U_{\text{L,B}}(t) \rangle}{U_{\text{L,B}}^* U_{\text{L,B}}}. \tag{2.19}$$

Then, with the same equation, she finds the cross-correlation coefficient between $U_{\text{L,B}}^*(t)$ and $U_{\text{H,B}}(t)$. If the former result yields the higher cross-correlation, then Eve has determined that Hypothesis (iii) is correct; otherwise, Hypothesis (iv) is valid.

*2.3.1.2   Unilateral Knowledge*

Eve has partial knowledge of only Alice's noises, that is, the noise generator outputs of Alice's resistors, $U_{\text{L,A}}(t)$ and $U_{\text{H,A}}(t)$. Bob's generator voltages are completely unknown to her. The attack methods described in Section 2.3.1.1 work even here with a minor modification as described below.

2.3.1.2.1   *Cross-correlations between Alice's/Bob's and Eve's channel voltages, currents and power*

Eve generates two independent "dummy" thermal noises to substitute for Bob's unknown noise voltages $U_{\text{H,B}}(t)$ and $U_{\text{L,B}}(t)$. Then, she uses the same protocol as the bilateral case (see Section 2.3.1.1): she measures $U_{\text{w}}(t)$ and $I_{\text{w}}(t)$, determines $P_{\text{w}}(t)$ from Equation (2.1), and uses Equations (2.13)-(2.15) to find the cross-correlation between her four simulated $U_{\text{w}}(t)$, $I_{\text{w}}(t)$, and $P_{\text{w}}(t)$, and the measured $U_{\text{w}}(t)$, $I_{\text{w}}(t)$, and $P_{\text{w}}(t)$.

2.3.1.2.2   *Cross-correlations between Alice's and Eve's voltage sources*

With $U_{\text{L,A}}(t)$ and $U_{\text{H,A}}(t)$ partially known, Eve uses a protocol corresponding to the bilateral case (see Section 2.3.1.1): she measures $U_{\text{w}}(t)$ and $I_{\text{w}}(t)$ [41, 42]. Then, from $I_{\text{w}}(t)$, she calculates the hypothetical voltage drops on Alice's possible choice of resistances $R_{\text{H}}$ and $R_{\text{L}}$. With these data, she tests two hypotheses:

**Hypothesis (i):** Alice has chosen $R_{\text{L}}$

**Hypothesis (ii):** Alice has chosen $R_{\text{H}}$.

Note that, in the bilateral case, she had knowledge of Bob's noises, thus she could form four hypotheses. In the unilateral case, she has knowledge of only Alice's generator voltages, therefore she can form only two hypotheses.

With Hypothesis (i), Eve uses Equation (2.16) to find $U_{L,A}^*(t)$. Then, she uses Equation (2.17) to find the cross-correlation between $U_{L,A}(t)$ and $U_{L,A}^*(t)$, and then the cross-correlation between $U_{L,A}^*(t)$ and $U_{H,A}(t)$. If the former result yields the higher cross-correlation, then Eve has determined that Hypothesis (i) is correct; otherwise, Hypothesis (ii) is valid.

The extra step that is needed to add to the protocol described in Section 2.3.1.1.2 is as follows: with Alice's chosen resistance known, Eve does the same exact KLJN protocol as Alice see Section 1.2): she evaluates the measured mean-square voltage on the wire over the whole bit exchange period. From that value, by using Equation (1.5), she evaluates the bit situation (see Figure 1.3). Thus she has cracked the KLJN scheme [41].

### 2.3.2 Demonstration

The noise generation is the same as in Sections 2.2.2.1 and 2.2.2.2.

*2.3.2.1 Attack demonstration when Eve knows both noises (bilateral attacks)*

*2.3.2.1.1 Bilateral attack demonstration utilizing cross-correlations between Alice's/Bob's and Eve's wire voltages, currents and powers*

Computer simulations were executed with MATLAB, and each simulation was averaged over 1000 runs, see Tables 2.1-2.4. During the attacks, the sample size (number of time steps) is 1000, and the KLJN system is kept in the LH state. A realization of the wire voltage, current, and power, $U_w(t)$, $I_w(t)$, and $P_w(t)$, under the LH condition over 100 milliseconds is displayed in Figure 2.17.

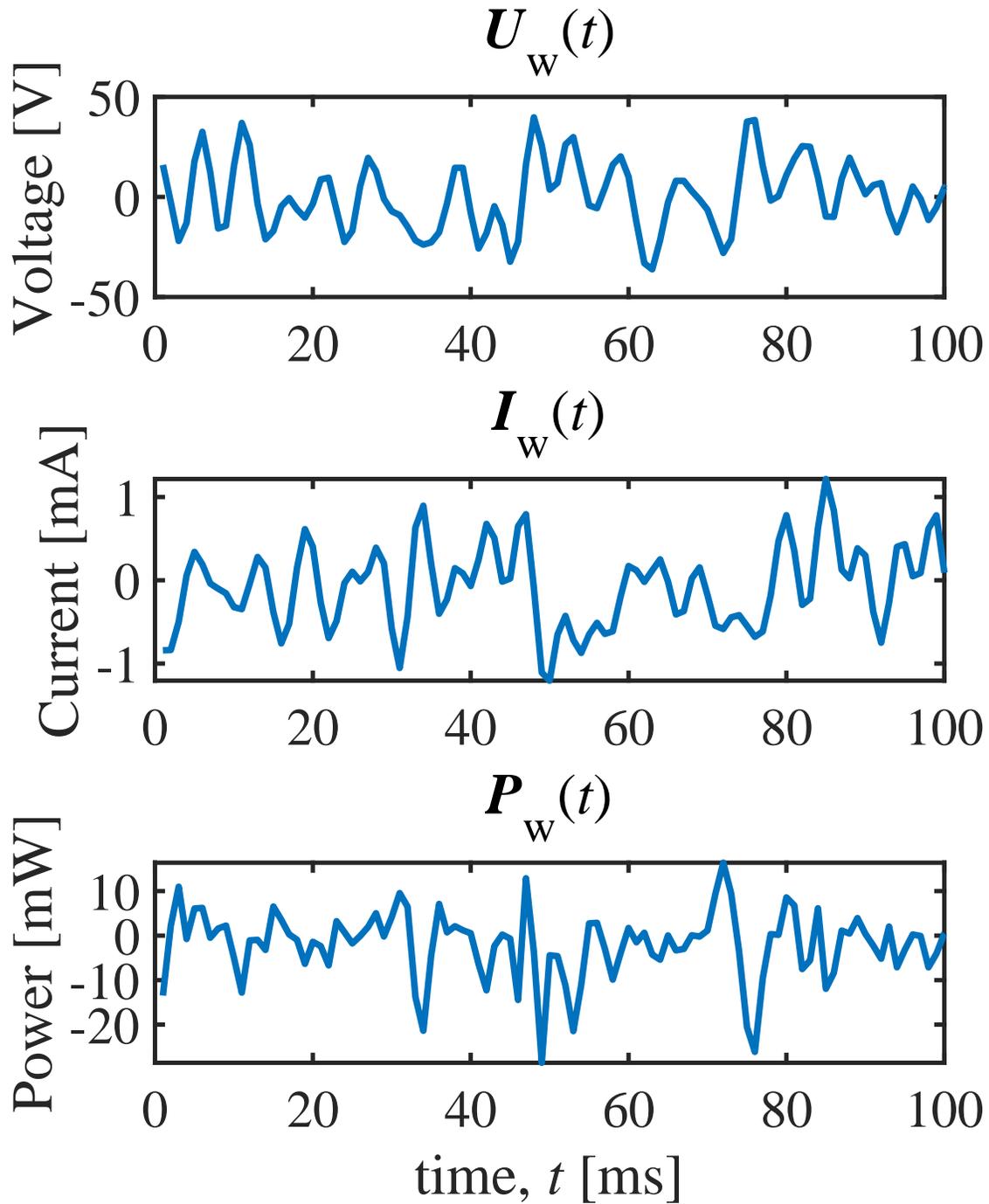

Figure 2.17: A realization of $U_w(t)$, $I_w(t)$, (see Figure 1.2) and $P_w(t)$ (see Equation (2.1)) for the LH situation displayed over 100 milliseconds [41]. Eve measures and records these data.

A visual example of the correlations between Eve's *measured* channel voltage, $U_w(t)$, and her four *simulated* channel voltages, $U_{\text{HH}}(t)$, $U_{\text{LL}}(t)$, $U_{\text{HL}}(t)$, and $U_{\text{LH}}(t)$, at $M = 1$ (see Section 2.2.2.2), is shown in Figure 2.18. The highest correlation is the LH case, thus Eve guesses that the secure resistance situation is LH.

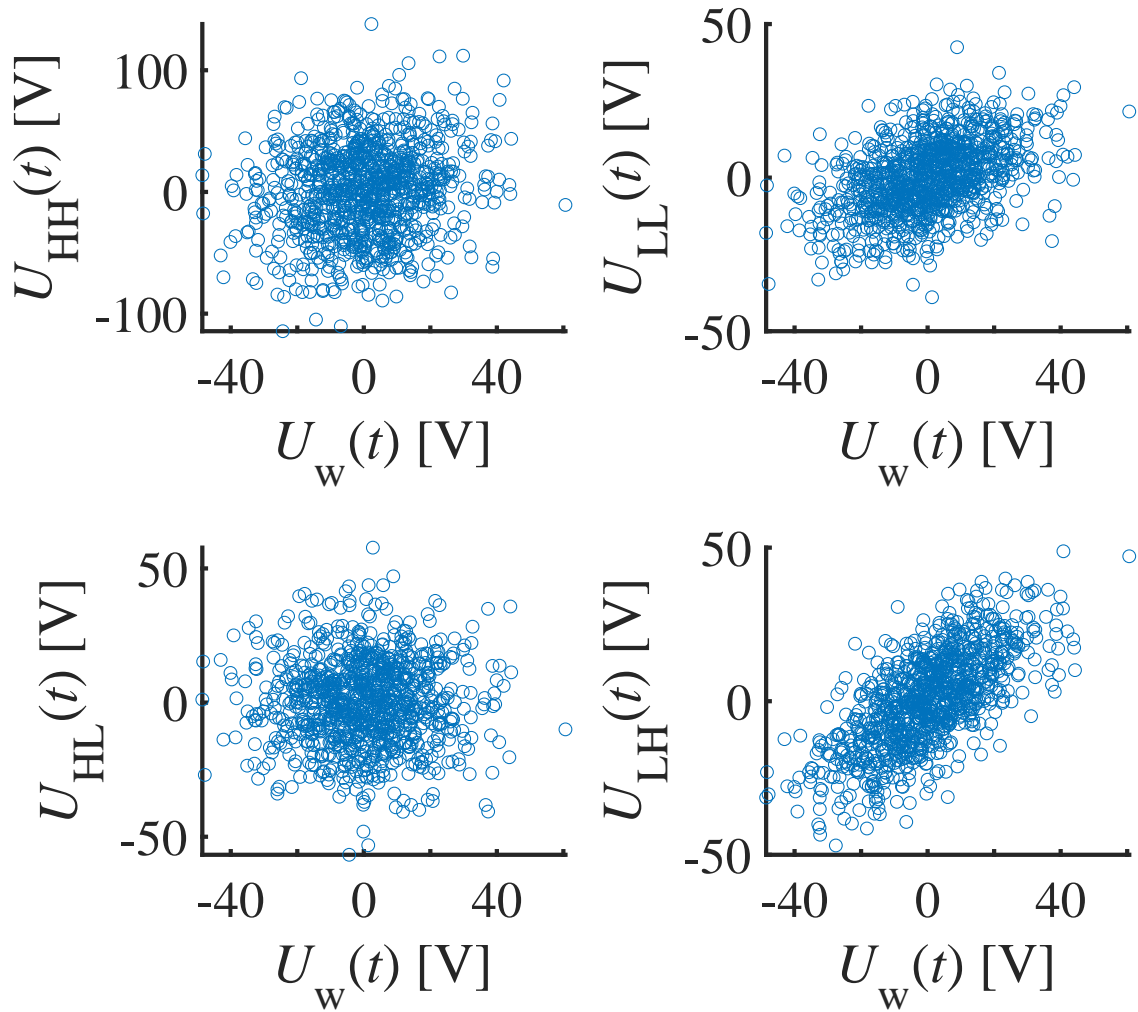

Figure 2.18: Correlation between Eve's *measured* channel voltage, $U_{\text{w}}(t)$, and her four *simulated* channel voltages, $U_{\text{HH}}(t)$, $U_{\text{LL}}(t)$, $U_{\text{HL}}(t)$, and $U_{\text{LH}}(t)$ at $M = 1$.

The results of the attack are shown in Table 2.1. The HL situation of Eve has zero cross-correlation with Eve's probe signals because, in this case, Alice/Bob and Eve do not have a common noise. In all the other cases, they have at least one mutual noise. In the LH attack situation, both noises are common, thus they yield the highest cross-correlation with Eve's LH probe, therefore Eve guesses that the secure resistance situation is LH.

Using the correlations between the voltages for the attack yielded the highest probability $p$ of successful guessing, and the correlations between the powers yielded the lowest one. This is because power is the product of voltage and current (see Equation (2.1)), and the inaccuracies in both the voltage and current correlations affect the power correlations.

2.3.2.1.2 *Bilateral attack demonstration utilizing cross-correlations among the voltage sources*

Eve uses Equation (2.16) to find $U^*_{L,A}(t)$, Equation (2.17) to find the cross-correlation between $U^*_{L,A}(t)$ and $U_{L,A}(t)$, Equation (2.18) to find $U^*_{L,B}(t)$, and Equation (2.19) to find the cross-correlation between $U^*_{L,B}(t)$ and $U_{L,B}(t)$.

A visual example of the correlations between (a) $U^*_{L,A}(t)$ and $U_{L,A}(t)$, $U^*_{L,A}(t)$ and $U_{H,A}(t)$, and (b) $U^*_{L,B}(t)$ and $U_{L,B}(t)$, $U^*_{L,B}(t)$ and $U_{H,B}(t)$, is shown in Figure 2.19. The highest correlation on Alice's side is with $U^*_{L,A}(t)$, and the highest correlation on Bob's side is with $U_{H,B}(t)$, thus Eve guesses that Alice has $R_L$ and Bob has $R_H$.

Table 2.1: Simulation of the average cross-correlation coefficient, $CCC$ (see Equations (2.13)-(2.15)), Eve's average $p$ of correctly guessing the LH bit situations, and standard deviation $\sigma$ at varying multipliers $M$ (see Section 2.2.2.2). As $M$ increases, which implies increasing differences between the noises of Alice/Bob and Eve's probing noises, the cross-correlation decreases. Yet, in each case of the present situation, the LH case yields the highest correlation. Thus, Eve guesses that LH is the secure bit situation. The correlations $CCC_\mathrm{u}$ between the voltages yielded $p_\mathrm{u}$, which had the highest probability $p$ of successful guessing, and the correlations $CCC_\mathrm{p}$ between the powers yielded the lowest $p$. This is because power is the product of voltage and current (see Equation (2.1)), and the inaccuracies in both the voltage and current correlations affect the power correlations.

| Eve's probing bit | $M$ | $CCC_\mathrm{u}$ | $p_\mathrm{u}$ | $\sigma_\mathrm{u}$ | $CCC_\mathrm{i}$ | $p_\mathrm{i}$ | $\sigma_\mathrm{i}$ | $CCC_\mathrm{p}$ | $p_\mathrm{p}$ | $\sigma_\mathrm{p}$ |
|---|---|---|---|---|---|---|---|---|---|---|
| HH | 0 | 0.21314 | | | 0.67455 | | | 0.28482 | | |
| LL | | 0.67377 | | | 0.21317 | | | 0.28502 | | |
| HL | | -0.00118 | | | 0.00128 | | | -0.00126 | | |
| LH | | 1 | 1 | 0 | 1 | 1 | 0 | 1 | 1 | 0 |
| HH | 0.1 | 0.21087 | | | 0.67048 | | | 0.28198 | | |
| LL | | 0.67090 | | | 0.21326 | | | 0.28573 | | |
| HL | | -0.00030 | | | -0.00007 | | | 0.00095 | | |
| LH | | 0.99504 | 1 | 0 | 0.99505 | 1 | 0 | 0.99005 | 1 | 0 |
| HH | 0.5 | 0.18975 | | | 0.60201 | | | 0.23101 | | |
| LL | | 0.60349 | | | 0.19047 | | | 0.23089 | | |
| HL | | -0.00023 | | | -0.00090 | | | -0.00114 | | |
| LH | | 0.89457 | 1 | 0 | 0.89441 | 1 | 0 | 0.79926 | 1 | 0 |
| HH | 1 | 0.15004 | | | 0.47631 | | | 0.14297 | | |
| LL | | 0.47705 | | | 0.15241 | | | 0.14315 | | |
| HL | | 0.00082 | | | 0.00074 | | | 0.00061 | | |
| LH | | 0.70664 | 1 | 0 | 0.70667 | 1 | 0 | 0.49926 | 1 | 0 |
| HH | 1.5 | 0.11814 | | | 0.37380 | | | 0.08931 | | |
| LL | | 0.37390 | | | 0.11798 | | | 0.08669 | | |
| HL | | 0.00022 | | | -0.00024 | | | -0.00141 | | |
| LH | | 0.55434 | 1 | 0 | 0.55473 | 1 | 0 | 0.30762 | 1 | 0 |
| HH | 5 | 0.04021 | | | 0.13280 | | | 0.01158 | | |
| LL | | 0.13142 | | | 0.03961 | | | 0.01000 | | |
| HL | | -0.00069 | | | 0.00087 | | | -0.00052 | | |
| LH | | 0.19490 | 1 | 0 | 0.19525 | 0.996 | 0.0144 | 0.03830 | 0.756 | 0.0144 |
| HH | 10 | 0.02200 | | | 0.06581 | | | 0.00266 | | |
| LL | | 0.06692 | | | 0.02130 | | | 0.00525 | | |
| HL | | 0.00064 | | | -0.00102 | | | 0.00090 | | |
| LH | | 0.09905 | 0.977 | 0.0034 | 0.09878 | 0.907 | 0.0111 | 0.01244 | 0.613 | 0.0209 |

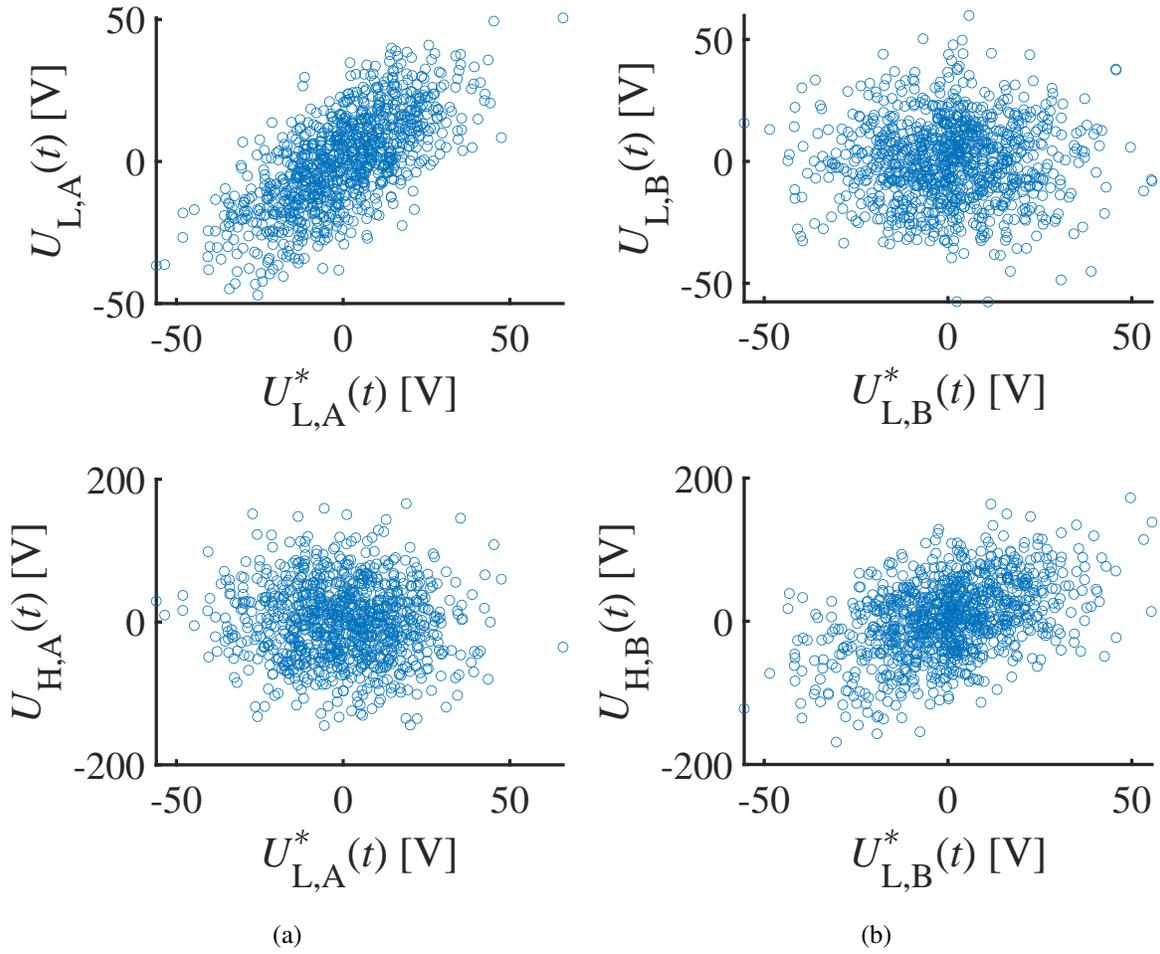

Figure 2.19: Correlation between (a) $U^*_{L,A}(t)$ vs. $U_{L,A}(t)$ and $U_{H,A}(t)$, and (b) $U^*_{L,B}(t)$ vs. $U_{L,B}(t)$ and $U_{H,B}(t)$. The highest correlation on Alice's side is with $U^*_{L,A}(t)$, and the highest correlation on Bob's side is with $U_{H,B}(t)$, thus Eve guesses that Alice has $R_L$ and Bob has $R_H$.

A simulation results of the cross correlation for each bit case is displayed in Table 2.2. As the multiplier $M$ in the superposition increases (see Section 2.2.2.2), the correlations decrease. Yet, in each case at the present (LH) situation, the highest correlation ended up being the LH case, thus Eve has decided that LH is the secure resistance situation.

Table 2.2: Simulation of the cross-correlation coefficients $CCC$ at the attack described in Section 2.3.1.1.2 (see Equations (2.17) and (2.19)), Eve's average correct-guessing probability $p$, and standard deviation $\sigma$ at varying multipliers $M$ (see Section 2.2.2.2). As $M$ increases, which implies increasing differences between the noises of Alice/Bob and Eve's probing noises, the cross-correlation decreases. Yet, in each case of the present situation, the $R_L$ hypothesis yields the highest cross-correlation with Alice's noise. Thus, Eve guesses that Alice has $R_L$. The same procedure done for Bob's noise results in the highest cross-correlation for the $R_H$ hypothesis at Bob's side.

| $M$ | $CCC_{LA}(R_L)$ | $CCC_{LA}(R_H)$ | $CCC_{LB}(R_L)$ | $CCC_{LB}(R_H)$ | $p$ | $\sigma$ |
|---|---|---|---|---|---|---|
| 0   | 1       | -0.03331 | -0.00092 | 0.57727 | 1     | 0      |
| 0.1 | 0.99504 | 0.00948  | 0.00031  | 0.55856 | 1     | 0      |
| 0.5 | 0.89451 | 0.01440  | 0.00063  | 0.51939 | 1     | 0      |
| 1   | 0.70682 | -0.05075 | 0.00140  | 0.40084 | 1     | 0      |
| 1.5 | 0.55430 | -0.02511 | 0.00004  | 0.31684 | 1     | 0      |
| 5   | 0.19496 | -0.03176 | -0.00088 | 0.14445 | 0.995 | 0.0022 |
| 10  | 0.09949 | -0.02127 | -0.00009 | 0.01611 | 0.894 | 0.0095 |

*2.3.2.2 Attack demonstration when Eve knows only one of the sources (unilateral attacks)*

*2.3.2.2.1 Unilateral attack demonstration utilizing cross-correlations between Alice's/Bob's and Eve's wire voltages, currents and powers*

Like the bilateral results (see Section 2.3.2.1.1), the HL situation of Eve has zero cross-correlation with Eve's probe signals because, in this case, Alice/Bob, and Eve do not have a common noise. In all the other cases, they have at least one mutual noise. In the LH probe attack, both noises are common, thus they yield the highest cross-correlation with Eve's LH probe, therefore Eve guesses that the secure resistance situation is LH, see Table 2.2.

Using the correlations between the voltages for the attack yielded the highest probability $p$ of successful guessing, and the correlations between the powers yielded the lowest one. This is because power is the product of voltage and current (see Equation (2.1)), and the inaccuracies in both the voltage and current correlations affect the power correlations.

Finally, Eve evaluates the measured mean-square voltage on the wire over the bit exchange period. From that value, by using Equation (1.5), she evaluates the parallel resultant $R_P$ of the resistances of Alice and Bob. From $R_\text{P}$ and $R_\text{A}$, she can calculate $R_\text{B}$. In this particular case, from the mean-square voltage Eve will learn that the actual situation is LH thus Bob has chosen $R_\text{H}$ because Alice has $R_\text{L}$ [41].

Table 2.3: Simulation of the cross-correlation coefficient, $CCC$ (see Equations (2.13)-(2.15)), Eve's average $p$ of correctly guessing the LH bit situations, and standard deviation $\sigma$ at varying multipliers $M$ (see Section 2.2.2.2). As $M$ increases, which implies increasing differences between the noises of Alice/Bob and Eve's probing noises, the cross-correlation decreases. Yet, in each case of the present situation, the LH case yields the highest correlation. Thus, Eve guesses that LH is the secure bit situation. The correlations $CCC_\text{u}$ between the voltages yielded $p_\text{u}$, which had the highest probability $p$ of successful guessing, and the correlations $CCC_\text{p}$ between the powers yielded the lowest $p$. This is because power is the product of voltage and current (see Equation (2.1)), and the inaccuracies in both the voltage and current correlations affect the power correlations.

| Eve's probing bit | $M$ | $CCC_\text{u}$ | $p_\text{u}$ | $\sigma_\text{u}$ | $CCC_\text{i}$ | $p_\text{i}$ | $\sigma_\text{i}$ | $CCC_\text{p}$ | $p_\text{p}$ | $\sigma_\text{p}$ |
|---|---|---|---|---|---|---|---|---|---|---|
| HH | 0 | -0.00166 | | | 0.00018 | | | -0.00055 | | |
| LL | | 0.67432 | | | 0.08975 | | | 0.16311 | | |
| HL | | 0.00055 | | | 0.00084 | | | -0.00193 | | |
| LH | | 0.90898 | 1 | 0 | 0.21237 | 0.995 | 0.0016 | 0.28506 | 0.982 | 0.0042 |
| HH | 0.1 | -0.00027 | | | -0.00048 | | | 0.00198 | | |
| LL | | 0.67119 | | | 0.09030 | | | 0.16500 | | |
| HL | | 0.00051 | | | -0.00006 | | | 0.00048 | | |
| LH | | 0.90462 | 1 | 0 | 0.21337 | 0.995 | 0.0026 | 0.28580 | 0.980 | 0.0044 |
| HH | 0.5 | 0.00022 | | | 0.00049 | | | 0.00224 | | |
| LL | | 0.60249 | | | 0.08149 | | | 0.13370 | | |
| HL | | -0.00168 | | | -0.00165 | | | 0.00062 | | |
| LH | | 0.81357 | 1 | 0 | 0.18865 | 0.983 | 0.0034 | 0.23146 | 0.962 | 0.0049 |
| HH | 1 | -0.00081 | | | 0.00079 | | | -0.0000 | | |
| LL | | 0.47610 | | | 0.06526 | | | 0.08290 | | |
| HL | | -0.00029 | | | 0.00009 | | | 0.00092 | | |
| LH | | 0.64237 | 1 | 0 | 0.15093 | 0.929 | 0.0132 | 0.14373 | 0.926 | 0.0062 |
| HH | 1.5 | -0.00046 | | | -0.00097 | | | 0.00001 | | |
| LL | | 0.37480 | | | 0.04921 | | | 0.04924 | | |
| HL | | 0.00159 | | | 0.00003 | | | -0.00016 | | |
| LH | | 0.50383 | 1 | 0 | 0.11863 | 0.859 | 0.0133 | 0.08675 | 0.827 | 0.0116 |
| HH | 5 | -0.00018 | | | 0.00094 | | | 0.00088 | | |
| LL | | 0.13188 | | | 0.01730 | | | 0.00728 | | |
| HL | | 0.00010 | | | 0.00168 | | | -0.00036 | | |
| LH | | 0.17768 | 1 | 0 | 0.04152 | 0.637 | 0.0157 | 0.01052 | 0.561 | 0.0143 |
| HH | 10 | 0.00058 | | | -0.00017 | | | -0.00088 | | |
| LL | | 0.06566 | | | 0.00989 | | | 0.00180 | | |
| HL | | -0.00108 | | | 0.00121 | | | -0.00096 | | |
| LH | | 0.08992 | 0.978 | 0.0055 | 0.02084 | 0.604 | 0.0134 | 0.00307 | 0.523 | 0.0170 |

Alternatively, since she knows Alice's resistance, she can use the mean-square voltage to determine the secure bit situation and thus crack the KLJN scheme.

2.3.2.2.2 *Unilateral attack demonstration utilizing cross-correlations among the voltage sources*

Eve uses (Equation (2.16)) to find $U^*_{\text{L,A}}(t)$ and (Equation (2.17)) to find the cross-correlation between $U^*_{\text{L,A}}(t)$ and $U_{\text{L,A}}(t)$. A simulation results of the cross correlation for each bit case is displayed in Table 2.4. As the multiplier $M$ increases (see Section 2.2.2.2), the correlation decreases, yet the highest correlation always ended up being the LH case, thus Eve has decided that LH is the secure resistance situation.

Table 2.4: A realization of the cross-correlation coefficient, $CCC$ (see Equation (2.17)), Eve's average correct-guessing probability $p$, and standard deviation $\sigma$ at varying multipliers $M$ (see Section 2.2.2.2). As $M$ increases, which implies increasing differences between the noises of Alice/Bob and Eve's probing noises, the cross-correlation decreases. Yet, the $R_\text{L}$ case yields the highest correlation. Thus, Eve guesses that Alice has $R_\text{L}$.

| $M$ | $CCC_\text{LA}(R_\text{L})$ | $CCC_\text{LA}(R_\text{H})$ | $p$ | $\sigma$ |
|---|---|---|---|---|
| 0 | 1 | -0.03331 | 1 | 0 |
| 0.1 | 0.99504 | 0.00948 | 1 | 0 |
| 0.5 | 0.89451 | 0.01440 | 1 | 0 |
| 1 | 0.70682 | -0.05075 | 1 | 0 |
| 1.5 | 0.55430 | -0.02511 | 1 | 0 |
| 5 | 0.19496 | -0.03176 | 1 | 0 |
| 10 | 0.09949 | -0.02127 | 0.989 | 0.0022 |

Finally, Eve evaluates the measured mean-square voltage on the wire over the bit exchange period. From that value, by using Equation (1.5), she evaluates the parallel resultant $R_\text{P}$ of the resistances of Alice and Bob. From $R_\text{P}$ and $R_\text{A}$, she can calculate $R_\text{B}$. In this particular case, from the mean-square voltage, Eve will learn that the actual situation is LH, thus Bob has chosen $R_\text{H}$ because Alice has $R_\text{L}$ [41].

### 2.3.3 Transition

This concludes the RNG attacks presented in this dissertation. As a take-home message, the security of the KLJN scheme requires the RNG outputs to appear truly random for Eve. Now, we move onto the zero-crossing attack.

# 3. ZERO-CROSSING ATTACK AGAINST THE KIRCHHOFF-LAW-JOHNSON-NOISE SECURE KEY EXCHANGER[4]

## 3.1 Security in Thermal Equilibrium

In the KLJN protocol, the net power flow $\langle P_\text{w}(t) \rangle$ between Alice and Bob (see Equation (2.1)) is zero because their resistors have the same (noise) temperature. The noise spectra of the voltage $U_\text{w}(t)$ and current $I_\text{w}(t)$ in the wire (see Figure 1.2), $S_\text{u}$ and $S_\text{i}$, respectively, are given by the Johnson formulas of thermal noise:

$$S_\text{u}(f) = 4kTR_\text{P}, \tag{3.1}$$

$$S_\text{i}(f) = \frac{4kT}{R_\text{S}}, \tag{3.2}$$

where $k$ is the Boltzmann constant, and $R_\text{P}$ and $R_\text{S}$ are the parallel and serial resultants of the connected resistors, respectively. In the HL and LH cases the resultants are:

$$R_{\text{P}_\text{LH}} = R_{\text{P}_\text{HL}} = \frac{R_\text{L} R_\text{H}}{R_\text{L} + R_\text{H}}, \tag{3.3}$$

$$R_{\text{S}_\text{LH}} = R_{\text{S}_\text{HL}} = R_\text{L} + R_\text{H}, \tag{3.4}$$

Equations (3.1)-(3.4) guarantee that the noise spectra and effective voltage and current values in the wire are identical in the LH and HL cases, in accordance with the perfect security requirement. In conclusion, the quantities that Eve can access with passive measurements satisfy the following equations that, together with Equations (3.3) and (3.4), form the pillars of security against passive

---

[4]Part of this chapter is reprinted with permission from C. Chamon, L. B. Kish, "Perspective–on the thermodynamics of perfect unconditional security," *Applied Physics Letters*, vol. 119, pp. 010501, 2021. Copyright 2021 by AIP Publishing.

attacks against the KLJN system:

$$U_{\text{LH}} = U_{\text{HL}}, \tag{3.5}$$

$$I_{\text{LH}} = I_{\text{HL}}, \tag{3.6}$$

$$P_{\text{LH}} = P_{\text{HL}} = 0 \tag{3.7}$$

where $U_{\text{LH}}$, $I_{\text{LH}}$, and $P_{\text{LH}}$ denote the RMS wire voltage, RMS wire current, and net power flow from Alice to Bob in the LH case, and $U_{\text{HL}}$, $I_{\text{HL}}$, and $P_{\text{HL}}$ denote the RMS wire voltage, RMS wire current, and net power flow from Alice to Bob in the HL case.

## 3.2 Security out of Equilibrium? The VMG-KLJN System

Vadai, Mingesz, and Gingl (VMG) have made an impressive generalization attempt [51], see Figure 3.1. They assumed that the four resistors are different and arbitrarily chosen (with some limitations [51]) and asked the question if the security can be maintained by a proper choice of different temperatures of these resistors.

In search for their solution, they used Equations (3.5) and (3.6) and removed Equation (3.7), the zero power flow condition.

VMG obtained the following solutions for the required mean-square thermal noise voltages of the thermal noise generators of the resistors, where the rms voltage $U_{\text{L,A}}$ of the resistor $R_{\text{L,A}}$ (see Figure 3.1) is freely chosen:

$$U_{\text{H,B}}^2 = U_{\text{L,A}}^2 \frac{R_{\text{L,B}}(R_{\text{H,A}} + R_{\text{H,B}}) - R_{\text{H,A}}R_{\text{H,B}} + R_{\text{H,B}}^2}{R_{\text{L,A}}^2 + R_{\text{L,B}}(R_{\text{L,A}} - R_{\text{H,A}}) - R_{\text{H,A}}R_{\text{L,A}}} = 4kT_{\text{H,B}}R_{\text{H,B}}\Delta f_{\text{B}}, \tag{3.8}$$

$$U_{\text{H,A}}^2 = U_{\text{L,A}}^2 \frac{R_{\text{L,B}}(R_{\text{H,A}} + R_{\text{H,B}}) + R_{\text{H,A}}R_{\text{H,B}} + R_{\text{H,A}}^2}{R_{\text{L,A}}^2 + R_{\text{L,B}}(R_{\text{L,A}} + R_{\text{H,B}}) + R_{\text{H,B}}R_{\text{L,A}}} = 4kT_{\text{H,A}}R_{\text{H,A}}\Delta f_{\text{B}}, \tag{3.9}$$

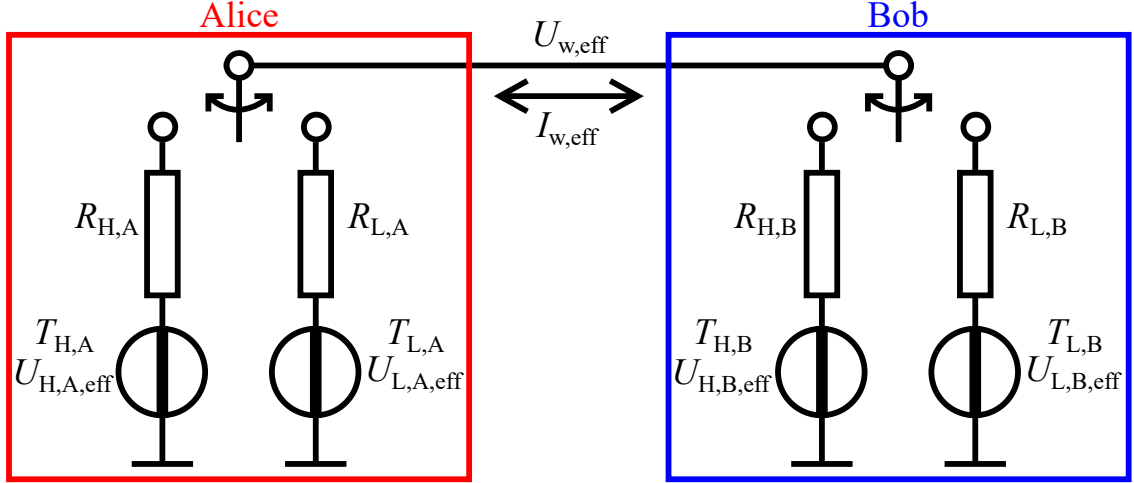

Figure 3.1: The core of the Vadai-Mingesz-Gingl (VMG-KLJN) secure key exchanger scheme. The four resistors are different and they can be freely chosen (though with some limitations because of certain unphysical solutions). One temperature is freely chosen. The other 3 temperatures depend on the resistor values and can be deducted by the VMG-equations (3.8)-(3.10), see Equations (3.11)-(3.13) below.

$$U_{L,B}^2 = U_{L,A}^2 \frac{R_{L,B}(R_{H,A} - R_{H,B}) - R_{H,A}R_{H,B} + R_{L,B}^2}{R_{L,A}^2 + R_{L,A}(R_{H,B} - R_{H,A}) - R_{H,A}R_{H,B}} = 4kT_{L,B}R_{L,B}\Delta f_B, \tag{3.10}$$

For more direct statistical physical comparison, we expanded [51] the VMG equations by introducing (on the right hand side) the temperatures of the resistors, where $\Delta f_B$ is the noise bandwidth of the generators, which is identical for all resistors, and the required temperatures of the resistors shown above are determined by the Johnson-Nyquist formula, see Equation (1.5). Thus, from Equations (3.8)-(3.10) and Equation (1.5), the temperatures are:

$$T_{H,B} = \frac{R_{L,A}}{R_{H,B}} T_{L,A} \frac{R_{L,B}(R_{H,A} + R_{H,B}) - R_{H,A}R_{H,B} + R_{H,B}^2}{R_{L,A}^2 + R_{L,B}(R_{L,A} - R_{H,A}) - R_{H,A}R_{L,A}}, \tag{3.11}$$

$$T_{H,A} = \frac{R_{L,A}}{R_{H,A}} T_{L,A} \frac{R_{L,B}(R_{H,A} + R_{H,B}) + R_{H,A}R_{H,B} + R_{H,A}^2}{R_{L,A}^2 + R_{L,B}(R_{L,A} + R_{H,B}) + R_{H,B}R_{L,A}}, \tag{3.12}$$

$$T_{\text{L,B}} = \frac{R_{\text{L,A}}}{R_{\text{L,B}}} T_{\text{L,A}} \frac{R_{\text{L,B}}(R_{\text{H,A}} - R_{\text{H,B}}) - R_{\text{H,A}} R_{\text{H,B}} + R_{\text{L,B}}^2}{R_{\text{L,A}}^2 + R_{\text{L,A}}(R_{\text{H,B}} - R_{\text{H,A}}) - R_{\text{H,A}} R_{\text{H,B}}}, \tag{3.13}$$

where $T_{\text{L,A}}$ is the temperature of resistor $R_{\text{L,A}}$.

The practical advantage of the VMG-KLJN scheme would appear with inexpensive versions of chip technology where resistance accuracy and its temperature stability are poor. However, concerning the fundamental physics aspects for security, there is a much more important question: What law of physics that guarantees the perfect security of the ideal system? In the case of the standard KLJN system, that is the *Second Law of Thermodynamics* (see Section 1.2). However, due to VMG's security claim at nonzero power flow, this explanation is seemingly irrelevant in the VMG-KLJN system.

### 3.2.1  The FCK1-VMG-KLJN System: Different Resistors but Still in Equilibrium

Recently, Ferdous, Chamon and Kish (FCK) [51] pointed out some excess information leak (compared to classical KLJN protocols) in the VMG-KLJN system under practical conditions. Among others, they proposed three modified VMG-KLJN versions for improvements. One of these schemes, the FCK1-VMG-KLJN scheme [51], is able to operate with four different resistors so that during each secure bit exchange period *the connected* resistor pair (one resistor at each side) is in thermal equilibrium, that is, the resistors in the pair have the same temperature. However, the two "secure-choice" resistor arrangements $R_{\text{H,A}} || R_{\text{L,B}}$ and $R_{\text{L,A}} || R_{\text{H,B}}$ must be at a different temperature, except in the original KLJN scheme where the two resistor pairs (of the HL and LH situations) are identical. (This is a minor security risk but is out of the topic of our present dissertation.)

The condition of zero power is that the geometrical means of the connected resistors in the LH and HL situations are equal [51]. In other words, when we choose three resistors freely, the fourth one is determined by the condition of zero power flow. For example, with $R_{\text{H,B}}$, $R_{\text{L,A}}$, and $R_{\text{H,A}}$ chosen, we get:

$$R_{\text{L,B}} = \frac{R_{\text{H,B}} R_{\text{L,A}}}{R_{\text{H,A}}}. \tag{3.14}$$

Due to the thermal equilibrium during a single bit exchange, the FCK1-VMG-KLJN system has a special role in the study of the non-equilibrium VMG-KLJN protocol in the following section.

In the next section, we answer the following question: Is it possible that there is a new, unknown attack that can extract information from the VMG-KLJN system while it is unable to do that with the standard KLJN scheme? If so, the VMG-KLJN arrangement would be just a modified KLJN scheme that is *distorted* for a special purpose (free resistor choice) while, as a compromise, its perfect security is given up. It would still have the same foundation of security, the Second Law, but in an imperfect way due to the nonideality introduced by the nonzero power flow. In the next two sections, we show that this is indeed the case.

### 3.3 ZERO-CROSSING ATTACK AGAINST THE VMG-KLJN SCHEME

The VMG-KLJN scheme seems to be perfectly secure at nonzero power flow because the voltage and the current are Gaussian processes and their mean-square values are identical in the LH and HL bit situations, even though the resistor and related mean-square voltage pairs are different. Therefore, even the power, which is the mean of their product (see Equation (2.1)), seems to carry no useful information for Eve. Gaussian processes are perfect information hiders.

Thus, we are exploring here a yet uncharted area: the statistics of the coincidence properties of the voltages at the two ends. Whenever the current is zero in the wire, the voltages at the two ends are equal. Let us *sample* the voltages on the wire at these coincidence points: then the voltage in the wire has the same value as that of the generators of Alice and Bob because the current is zero. For an intuitive start, imagine the situation when in the HL case the VMG voltage is very high at the H side and small at the L side, while in the LH case the voltages are similar (see Equations (3.8)-(3.10)). The Gaussian process is statistically confined to the order of the RMS value, thus these samples will be mostly confined to a fraction of the RMS value of the large noise at the H side of the HL situation, which is very different from what we have in the LH situation outlined above.

In this way, we have a heuristic hope that the mean-square values of these voltage samples will depend on the bit situations (HL or LH). Note, such an attack would not work against the original KLJN scheme, as there the HL and LH voltage and resistor pairs are identical.

Moreover, whenever the net power flow is zero due to thermal equilibrium, such as in the original KLJN scheme, the wire voltage and current are uncorrelated. Then their Gaussianity implies that sampling the voltage at the zero-crossing times of the current represents an independent sampling of the voltage. That means the mean-square wire voltage will be the nominal value for the HL/LH situation, thus there is no information there for Eve. This is another reason why the original KLJN scheme would be immune against such a zero-crossing attack. Moreover, it is an indication that the FCK1-VMG-KLJN system, where the power flow is also zero, is also immune against this new attack.

Below, we demonstrate by computer simulations that the intuitive expectation turns out to be valid, and the VMG-KLJN scheme is leaking information at nonzero power flow. In the next section, we also show that there is a new KLJN scheme that is secure against the attack even though the four resistors are different.

### 3.3.1 Computer Simulations/Verification of the Zero-Crossing Attack

During the noise generation, we used oversampling and interpolation to produce sufficiently smooth noises to emulate physical noise sources and to detect the zero-crossing current events with sufficient accuracy, see Figure 3.2.

An example for Alice's and Bob's noise voltages and channel current in the VMG-KLJN scheme, at $R_{H,A} = 46,416$ $\Omega$, $R_{L,A} = 278$ $\Omega$, $R_{H,B} = 278$ $\Omega$, $R_{L,B} = 100$ $\Omega$, $T_{H,A} = 8.0671 \times 10^{18}$ K, $T_{L,A} = 1.3033 \times 10^{17}$ K, $T_{H,B} = 6.2112 \times 10^{16}$ K, $T_{L,B} = 1.1694 \times 10^{17}$ K, and $\Delta f_B = 500$ Hz, is shown in Figure 3.2. The zero-crossing points of the channel current are the points where Alice's and Bob's noise voltages are equal. At the particular choice of resistances, in the LH case, Alice and Bob have similar noise voltage amplitudes, while in the HL case, Alice's noise voltage amplitudes are much larger than Bob's, thus the zero-crossing points are ultimately determined by Bob's noise voltage.

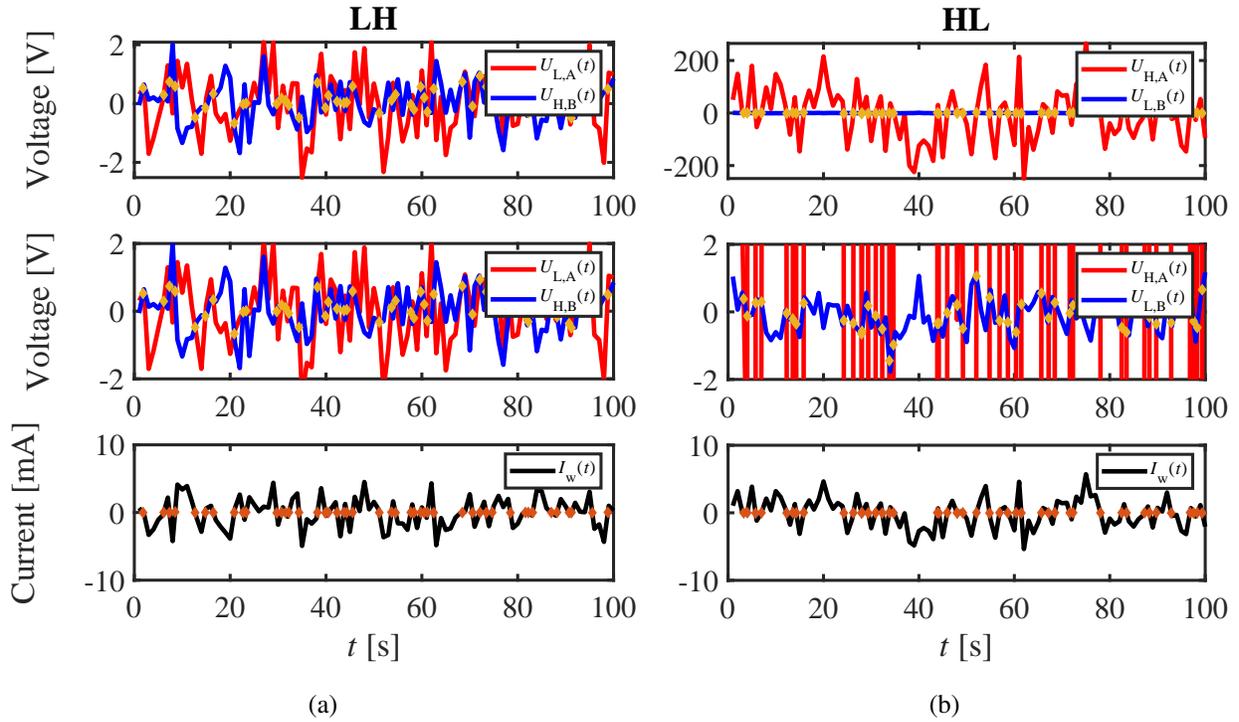

Figure 3.2: A realization of the instantaneous noise voltages of Alice (red) and Bob (blue) and the channel current (black) in the LH (a) and HL (b) cases for the VMG-KLJN scheme, at $R_{H,A} = 46,416\ \Omega$, $R_{L,A} = 278\ \Omega$, $R_{H,B} = 278\ \Omega$, $R_{L,B} = 100\ \Omega$, $T_{H,A} = 8.0671 \times 10^{18}$ K, $T_{L,A} = 1.3033 \times 10^{17}$ K, $T_{H,B} = 6.2112 \times 10^{16}$ K, $T_{L,B} = 1.1694 \times 10^{17}$ K, and $\Delta f_B = 500$ Hz. $U^2_{L,A} = 1$ V$^2$, $U^2_{H,B} = 0.477$ V$^2$, $U^2_{H,A} = 1.03 \times 10^4$ V$^2$, and $U^2_{L,B} = 0.323$ V$^2$. The points where the channel current $I_w(t)$ is zero, represented in orange, are the points where Alice's and Bob's noise voltages are equivalent, represented in yellow. In the LH case, Alice's noise voltage $U_{L,A}(t)$ is comparable to Bob's noise voltage $U_{H,B}(t)$, while in the HL case, Alice's noise voltage $U_{L,A}(t)$ is significantly larger than Bob's noise voltage $U_{H,B}(t)$, thus the points where Alice's and Bob's noise voltages are equal to each other are ultimately determined by the smaller noise amplitude. $U_{H,A} \gg U_{L,B}$, thus $U_{L,B}(t)$ looks like a straight line because of limited resolution in the figure. The middle subplot in (b) shows an enlarged scale to visualize crossing events while figure on the left is the same as above for comparison purposes.

The histograms of the mean-square channel voltages, currents, and zero-crossing points after 1,000 runs are shown in Figure 3.3 for the original KLJN scheme (a), the VMG-KLJN scheme (b), and the FCK1-VMG-KLJN scheme (c). The orange histograms represent the LH situation, whereas the blue histograms represent the HL situation. The red lines represent the expected (mean) value. The secure bit (LH and HL) mean-square voltages and currents are the same in all schemes, as it has been expected by the VMG-KLJN creators. However, in the LH and HL cases, the zero-crossing sampled mean-square voltages $U_{\text{w,zc}}^2$ are the same only in the original KLJN and FCK1-VMG-KLJN schemes, but markedly different in the VMG-KLJN scheme, indicating its cracked security.

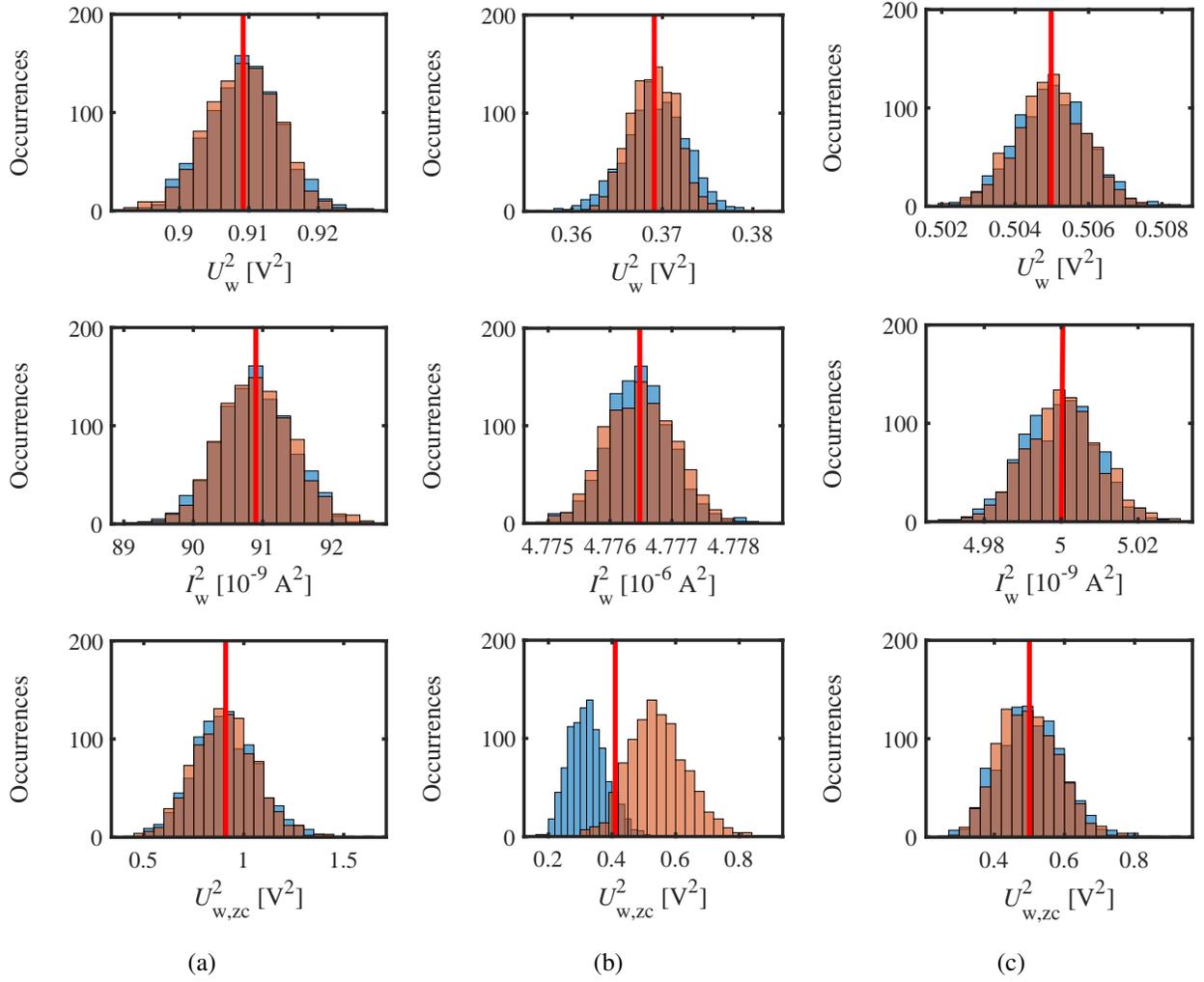

Figure 3.3: Histograms of the mean-square channel voltage $U_w^2$ (first row), current $I_w^2$ (second row), and zero-crossing points $U_{w,zc}^2$ (third row) for:
Column-(a) the original KLJN scheme at $R_{H,A} = R_{H,B} = 10$ kΩ and $R_{L,A} = R_{L,B} = 1$ kΩ,
Column-(b) the VMG-KLJN scheme at $R_{H,A} = 46.4$ kΩ, $R_{L,A} = 278$ Ω, $R_{H,B} = 278$ Ω, and $R_{L,B} = 100$ Ω, and
Column-(c) the FCK1-VMG-KLJN scheme at $R_{H,A} = 100$ kΩ, $R_{L,A} = 10$ kΩ, $R_{H,B} = 10$ kΩ, and $R_{L,B} = 1$ kΩ. The orange histograms represent the LH situation, and the blue histograms represent the HL situation. The red vertical lines represent the expected (mean) value. In all three schemes, $U_w^2$ and $I_w^2$ have the same LH and HL distributions (within statistical inaccuracy). In the original KLJN and FCK1-VMG-KLJN schemes, $U_{w,zc}^2$ has the same LH and HL distributions, in accordance with their perfect security. In the VMG-KLJN scheme, the distributions of the $U_{w,zc}^2$ values at the LH and HL cases are split, which indicates significant information leak.

Table 3.1 shows the mean-square voltage $U_w^2$, mean-square current, $I_w^2$, average power $\langle P_w(t) \rangle$,

and zero-crossing mean-square voltage $U_{w,zc}^2$ values for the original KLJN, three VMG-KLJN, and FCK1-VMG-KLJN representations.

In the original KLJN and FCK1-VMG-KLJN representations, the mean-square zero-crossing voltage approaches the channel voltage indicating a random, current-independent sampling. In the VMG-KLJN scheme, as the cross-correlation between the voltage and current increases (indicated also by the nonzero power flow), the mean-square zero-crossing voltage becomes more dispersed in the LH and HL cases.

Table 3.2 shows the statistical run for Eve's probability $p$ and its standard deviation $\sigma_p$ of guessing the correct bit. When the average power $\langle P_w(t) \rangle$ approaches zero, the $p$ value approaches 0.5 (thus the information leak converges zero) because the cross-correlation coefficient between the current and voltage also converges to zero.

Table 3.1: Results for the wire mean-square voltage $U_w^2$, mean-square current, $I_w^2$, average power $\langle P_w(t) \rangle$, and zero-crossing mean-square voltage $U_{w,zc}^2$ for the KLJN, three VMG-KLJN, and FCK1-VMG-KLJN schemes, where $R_A$ and $R_B$ represent Alice's and Bob's resistor choices, respectively. In the classical KLJN and FCK1-VMG-KLJN schemes, $U_{w,zc}^2$ approaches $U_w^2$. In the VMG-KLJN scheme, as $\langle P_w(t) \rangle$ increases, $U_{w,zc}^2$ becomes split in the LH and HL situations.

| Scheme | bit | $R_A$ [Ω] | $R_B$ [Ω] | $U_w^2$ [V] | $I_w^2$ [$10^{-6}$ A$^2$] | $\langle P_w(t) \rangle$ [$10^{-3}$ W] | $U_{w,zc}^2$ [V$^2$] |
|---|---|---|---|---|---|---|---|
| KLJN | LH | 1k | 10k | 0.908 | 0.091 | 0 | 0.907 |
|  | HL | 10k | 1k |  |  |  | 0.908 |
| VMG-KLJN | LH | 100 | 16.7k | 0.991 | 0.314 | 0.026 | 0.989 |
|  | HL | 16.7k | 278 |  |  |  | 1.009 |
|  | LH | 278 | 278 | 0.368 | 4.786 | 0.471 | 0.301 |
|  | HL | 46.4k | 100 |  |  |  | 0.576 |
|  | LH | 100 | 6k | 0.967 | 0.073 | 0.156 | 0.675 |
|  | HL | 360k | 2.2k |  |  |  | 0.845 |
| FCK-VMG-KLJN | LH | 10k | 10k | 0.500 | 0.005 | 0 | 0.498 |
|  | HL | 100k | 1k |  |  |  | 0.502 |

In conclusion, the computer simulations confirmed that the zero-crossing attack is an efficient passive attack against the general VMG scheme whenever the net power flow is not zero. The FCK1-VMG-KLJN protocol, which is the zero-power version of the scheme, is robust against this attack similarly to the original KLJN scheme.

Table 3.2: Statistical run for Eve's probability $p$ of guessing the correct bit from the zero-crossing attack on each scheme. When the average power $\langle P_{\text{w}}(t)\rangle$ approaches zero, the $p$ value approaches 0.5 (thus the information leak converges zero) because the cross-correlation coefficient between the current and voltage also converges to zero.

| Scheme | bit | $R_{\text{A}}$ [$\Omega$] | $R_{\text{B}}$ [$\Omega$] | $\langle P_{\text{w}}(t)\rangle$ [$10^{-3}$ W] | $p$ | $\sigma_p$ |
|---|---|---|---|---|---|---|
| KLJN | LH | 1k | 10k | 0 | 0.5002 | 0.0091 |
| | HL | 10k | 1k | | | |
| VMG-KLJN | LH | 100 | 16.7k | 0.026 | 0.5885 | 0.0022 |
| | HL | 16.7k | 278 | | | |
| | LH | 278 | 278 | 0.471 | 0.7006 | 0.0053 |
| | HL | 46.4k | 100 | | | |
| | LH | 100 | 6k | 0.156 | 0.6281 | 0.0021 |
| | HL | 360k | 2.2k | | | |
| FCK1-VMG-KLJN | LH | 10k | 10k | 0 | 0.5028 | 0.0091 |
| | HL | 100k | 1k | | | |

### 3.3.2 Transition

This concludes the zero-crossing attack presented in this dissertation. As a take-home message, thermal equilibrium is a requirement in the KLJN scheme. Now, we move onto the nonlinearity attack.

# 4. NONLINEARITY ATTACK AGAINST THE KIRCHHOFF-LAW-JOHNSON-NOISE (KLJN) SECURE KEY EXCHANGE PROTOCOL[5]

## 4.1 Nonlinearity

The noise generators of Alice and Bob have analog amplifiers as drivers. These have nonlinear characteristics [107]. We can model their output voltage by taking the Taylor Series approximation

$$U^*(t) = A \left[ U(t) + BU^2(t) + CU^3(t) + ... \right], \tag{4.1}$$

where $U^*(t)$ is the output voltage of the generator, $A$ is the linear amplification, $U(t)$ is the input noise voltage, and $B$ and $C$ are the second and third order nonlinearity coefficients, respectively.

Nonlinearity obviously distorts the amplitude distribution function and the Gaussianity of the noise sources. Vadai, Mingesz, and Gingl mathematically proved [59] that the KLJN scheme is secure only if the distribution of the noise voltages is Gaussian. Thus, nonlinearity is expected to cause information leak in these systems. It is an open question how much is this leak at practical conditions.

In this dissertation, we explore the effect of nonlinearity at the second order, third order, and a combination of the two orders. We also show that, as we decrease $T_{\text{eff}}$, the KLJN scheme approaches perfect security because the nonlinear components get negligibly small due to the reduced noise voltage.

## 4.2 The Nonlinearity Attack

The overview of the nonlinearity attack is shown in Figure 4.1. Alice's and Bob's key exchangers have a nonlinearity component to them, and Eve measures the power flow from Alice to Bob to guess the secure key bit situation.

---

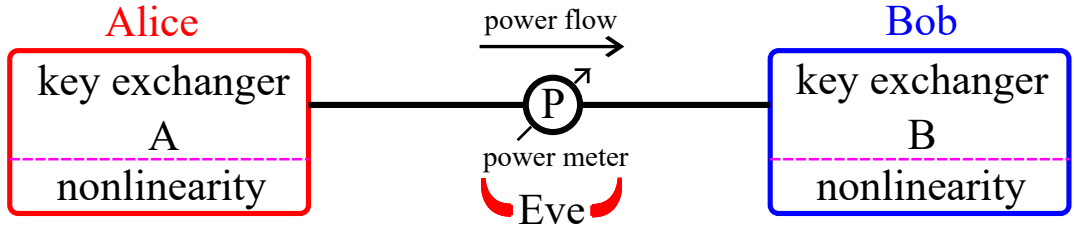

Figure 4.1: Overview of the nonlinearity attack. Alice's and Bob's key exchangers have a nonlinearity component to them, and Eve measures the power flow from Alice to Bob to guess the secure key bit situation.

For illustrative purposes, we use only the second and third order nonlinearities to account for the effects of the even and odd order nonlinearities. To quantify the nonlinearity, we use the total distortion, given by the sum of the normalized mean-square components:

$$\mathrm{TD} = \frac{\sqrt{\langle [BU^2(t)]^2 \rangle + \langle [CU^3(t)]^2 \rangle}}{\langle [U(t)]^2 \rangle}. \qquad (4.2)$$

Eve measures the channel voltage and current, $U_\mathrm{w}(t)$ and $I_\mathrm{w}(t)$ (see Figure 1.2) and calculates the net power flow from Alice to Bob (see Equation (2.1)),

$$\langle P_\mathrm{w}(t) \rangle = \langle I_\mathrm{w}(t) U_\mathrm{w}(t) \rangle, \qquad (4.3)$$

where the interpretation of voltage and current polarities are properly chosen for the direction of the power flow. Suppose the following protocol is publicly shared between Alice and Bob:

**(i)** If the net power flow is greater than zero, Eve surmises that HL is the secure bit situation;

**(ii)** If the net power flow is less than zero, Eve surmises that LH is the secure bit situation.

For example, in accordance with Equations (1.5) and (4.1) we conclude: In the case of positive nonlinear coefficients in Equation (4.1), the HL case means a higher mean-square voltage and a higher temperature at Alice's end, thus a positive power flow from Alice to Bob. If Eve extracts a key, she can test that key or its inverse. One of them will be the true key. (For example, with proper negative coefficients, HL can imply a negative power flow, which would lead to the inverse key. If Eve, in accordance with Kerckhoffs's principle, knows the nonlinear coefficient in Equation (4.1), the inverse operation with the key is not needed.)

## 4.3 Demonstration

Computer simulations with Matlab measure the information leak with practical nonlinearity parameters in Equation (4.1). The tests show a significant amount of information leak, even with small nonlinearity.

The protocol is as follows:

- For each bit exchange, Eve measures and evaluates the average power at the information channel $\langle P_w(t) \rangle$ (see Equation (4.3)).

- If the result is greater than zero, she guesses that HL is the secure bit situation;

- If the result is less than zero, she guesses that LH is the secure bit situation (see Section 4.2).

- The process above is independently repeated 1,000 times to obtain the statistics shown.

Out of the linear (Ideal) case, the investigated nonlinear situations are:

(a) case $D_2$ with second-order nonlinearity;

(b) case $D_3$ with third-order nonlinearity;

(c) case $D_{2,3}$ with third-order nonlinearity;

Figure 4.2 illustrates the IU scatterplots between the wire voltage and current for the Ideal (a), $D_2$ (b), $D_3$ (c), and $D_{2,3}$ (d) situations. The chosen parameters are $R_H = 100$ k$\Omega$, $R_L = 10$ k$\Omega$, $T_{\text{eff}} = 10^{18}$ K, and $\Delta f_B = 500$ Hz. At $D_2$, $B = 6 \times 10^{-3}$ and $C = 0$. At $D_3$, $B = 0$ and $C = 5 \times 10^{-5}$. At $D_{2,3}$, $B = 1 \times 10^{-6}$ and $C = 5 \times 10^{-5}$. The blue circles represent the HL case, whereas the orange crosses represent the LH case.

The HL and LH situations are statistically indistinguishable in the Ideal (linear) situation, indicating perfect security.

In the $D_2$ case, the HL arrangement has an upward dominance, while the LH has a downward tendency. In the $D_3$ and $D_{2,3}$ cases, the HL situation has a right-diagonal footprint, while the LH situation has a left-diagonal footprint. In conclusion, the nonlinear IU scatterplots indicate lack of security at the given conditions.

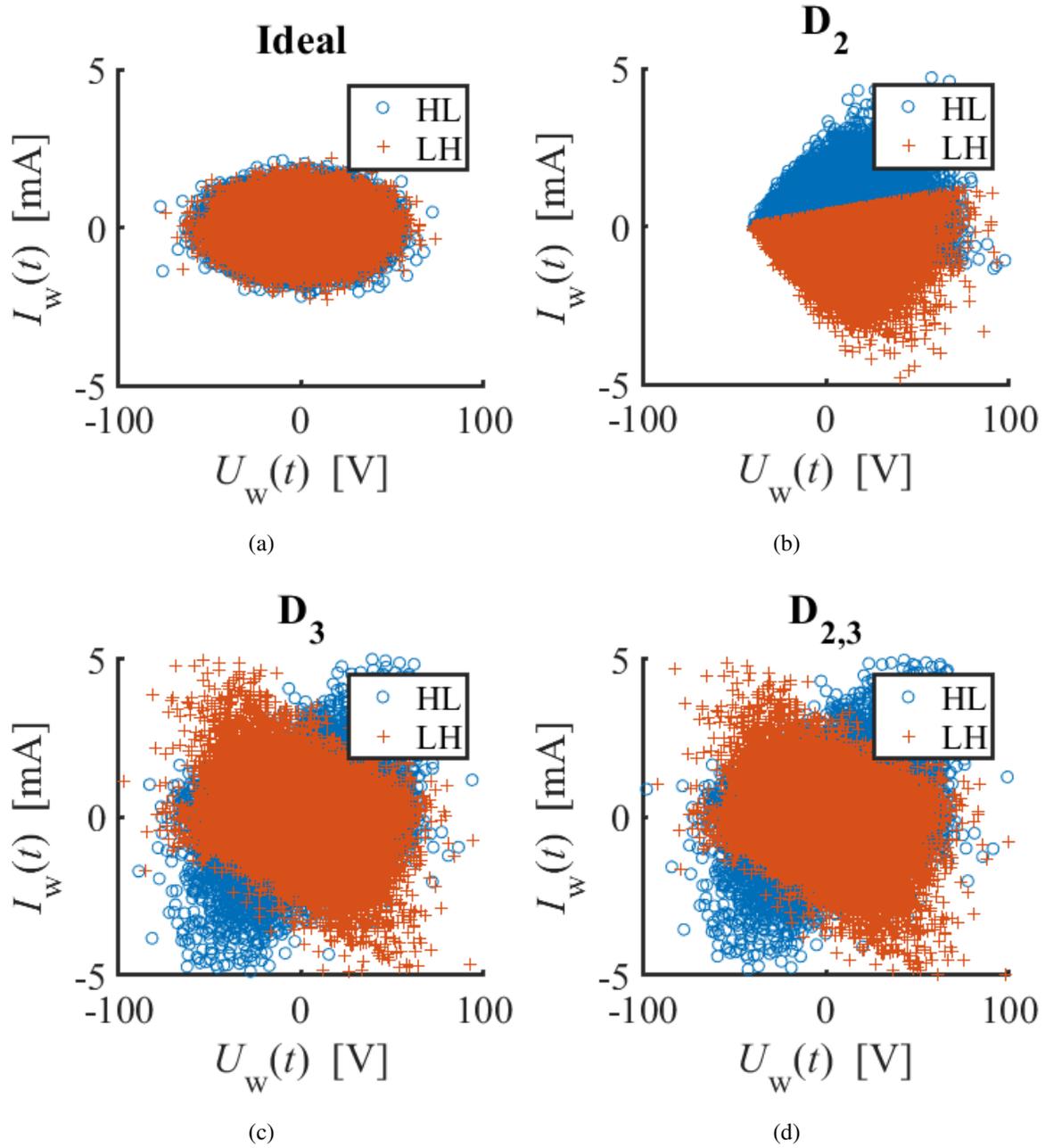

Figure 4.2: The IU scatterplots between the wire voltage and current for the ideal (a), $D_2$ (b), $D_3$ (c), and $D_{2,3}$ (d) situations. The parameters chosen are $R_H = 100$ kΩ, $R_L = 10$ kΩ, $T_{\text{eff}} = 10^{18}$ K, and $\Delta f_B = 500$ Hz. At $D_2$, $B = 6 \times 10^{-3}$ and $C = 0$. At $D_3$, $B = 0$ and $C = 5 \times 10^{-5}$. At $D_{2,3}$, $B = 1 \times 10^{-6}$ and $C = 5 \times 10^{-5}$. The blue circles represent the HL case, whereas the orange crosses represent the LH case. The HL and LH situations are statistically indistinguishable in the ideal situation. In the $D_2$ case, the HL arrangement has an upward dominance, while the LH has a downward tendency. In the $D_3$ and $D_{2,3}$ cases, the HL situation has a right-diagonal trajectory, while the LH has a left-diagonal trajectory.

Table 4.1 shows the statistical run for Eve's probability $p$ of correctly guessing the bit situations, and its standard deviation $\sigma$, for four different sample sizes (time steps) $\gamma$. For each nonlinearity situation, the $p$ value increases as $\gamma$ increases, as expected, due to the increasing accuracy of Eve's statistics.

Table 4.1: The statistical run for Eve's correct-guessing probability $p$ and its standard deviation $\sigma$ for four different sample sizes $\gamma$. For each nonlinearity situation, the $p$ value increases as $\gamma$ increases

| D | $\gamma$ | $p$ | $\sigma$ |
|---|---|---|---|
| 2 | 10 | 0.5502 | 0.0135 |
|   | 20 | 0.6172 | 0.0203 |
|   | 100 | 0.7498 | 0.0149 |
|   | 1000 | 0.9869 | 0.0042 |
| 3 | 10 | 0.5632 | 0.0159 |
|   | 20 | 0.5982 | 0.0140 |
|   | 100 | 0.7383 | 0.0126 |
|   | 1000 | 0.9831 | 0.0047 |
| 2,3 | 10 | 0.5761 | 0.0114 |
|   | 20 | 0.6106 | 0.0166 |
|   | 100 | 0.7434 | 0.0137 |
|   | 1000 | 0.9855 | 0.0037 |

Varying the effective temperature $T_{\text{eff}}$ resulted in varying the effective voltage on the wire $U_{\text{w}}$ (see Equation (1.5)). The statistical protocol with results shown in Table 4.1 was repeated for various effective temperatures.

Figure 4.3 illustrates Eve's correct-guessing probability $p$ (top) and Eve's bit error $\epsilon$ (bottom), given by

$$\epsilon = 1 - p, \tag{4.4}$$

with respect to the effective wire voltage $U_{\text{w}}$ for $\text{D}_2$ (a), $\text{D}_3$ (b), and $\text{D}_{2,3}$ (c) for all sample sizes $\gamma$. As $\gamma$ and $U_{\text{w}}$ (that is, the effective temperature) decrease, $p$ approaches perfect security.

Figure 4.4 shows $p$ and $\epsilon$ vs. $U_\text{w}$ at $\gamma = 1000$ for all the distortions. With the given parameters, convergence toward perfect security happened at $D_2$ before $D_3$ and $D_{2,3}$.

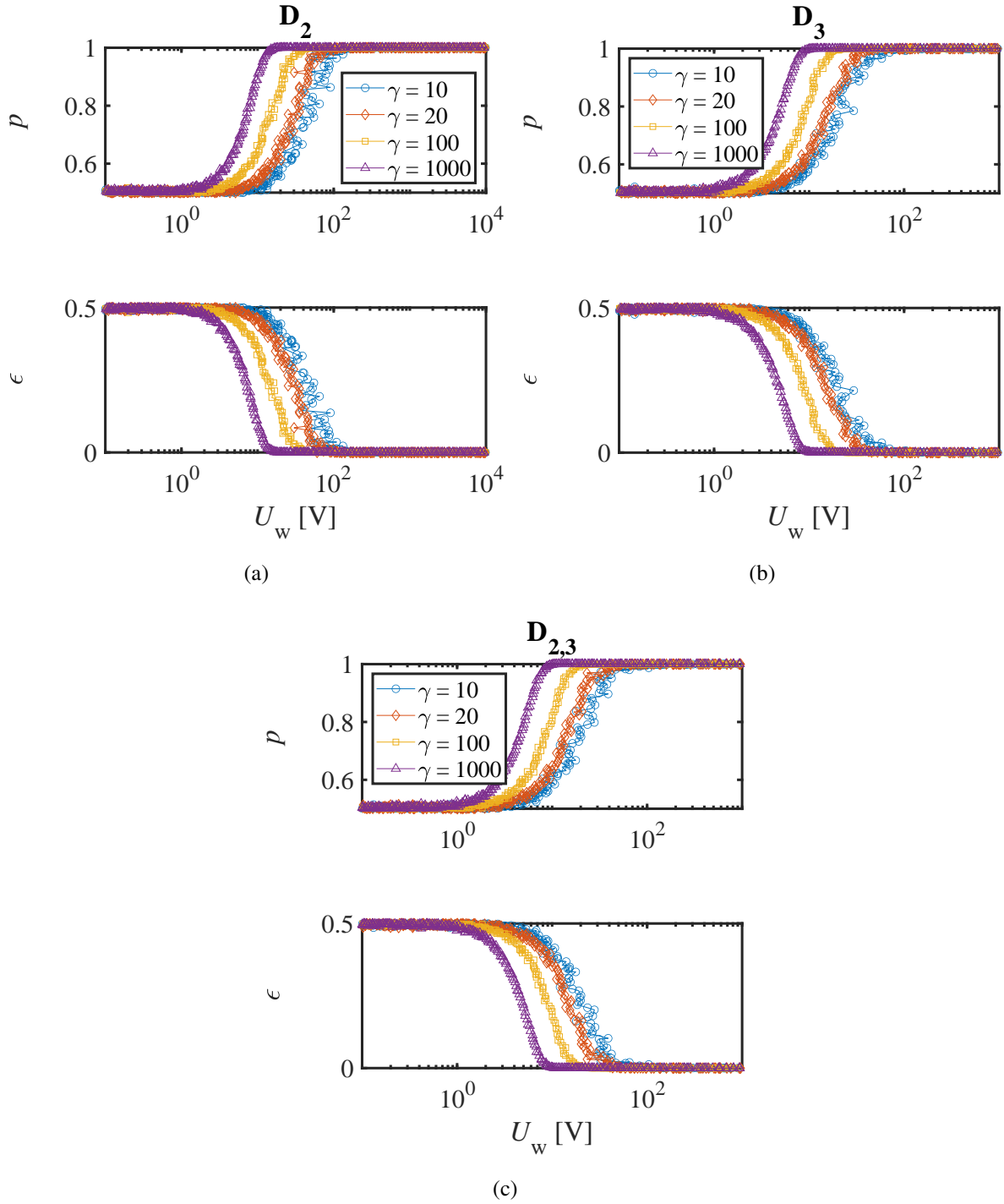

Figure 4.3: Eve's correct-bit-guessing probability $p$ (top) and Eve's bit error $\epsilon$ (bottom) with respect to the effective voltage $U_w$ for: $D_2$ (a), $D_3$ (b), and $D_{2,3}$ at $\gamma = 10$ (blue), $\gamma = 20$ (orange), $\gamma = 100$ (yellow), and $\gamma = 1000$ (purple). As $\gamma$ and $U_w$ (driven by the effective temperature) decrease, $p$ approaches perfect security.

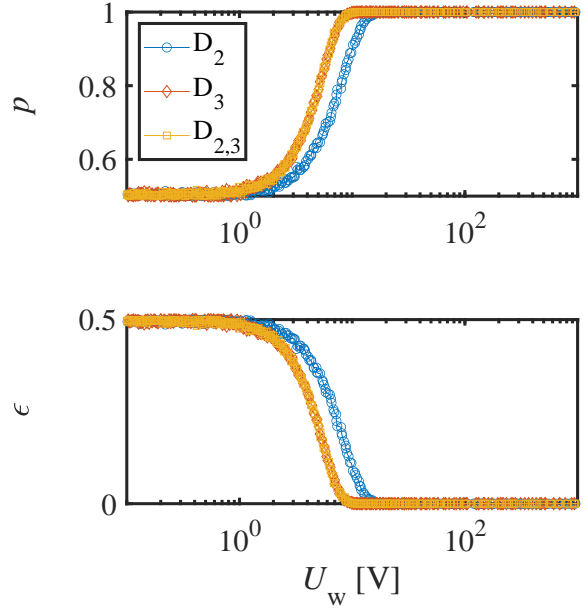

Figure 4.4: Eve's correct-bit-guessing probability $p$ (top) and Eve's bit error $\epsilon$ (bottom) with respect to the effective voltage $U_w$ at $\gamma = 1000$ for $D_2$, $D_3$, and $D_{2,3}$. $p$ increases and $\epsilon$ decreases as $U_w$ (driven by the effective temperature) increases. Convergence to perfect security happens at $D_2$ before $D_3$ and $D_{2,3}$.

Varying the effective temperature $T_{\text{eff}}$ also resulted in varying the effective current on the wire $I_w$. Figure 4.5 illustrates Eve's correct-guessing probability $p$ (top) and Eve's bit error $\epsilon$ (bottom, see Equation (4.4)), with respect to the effective wire current $I_w$ for $D_2$ (a), $D_3$ (b), and $D_{2,3}$ (c) for all sample sizes $\gamma$. As $\gamma$ and $I_w$ (that is the effective temperature) decrease, $p$ approaches perfect security.

Figure 4.6 shows $p$ and $\epsilon$ vs. $I_w$ at $\gamma = 1000$ for all the distortions. With the given parameters, convergence toward perfect security happened at $D_2$ before $D_3$ and $D_{2,3}$.

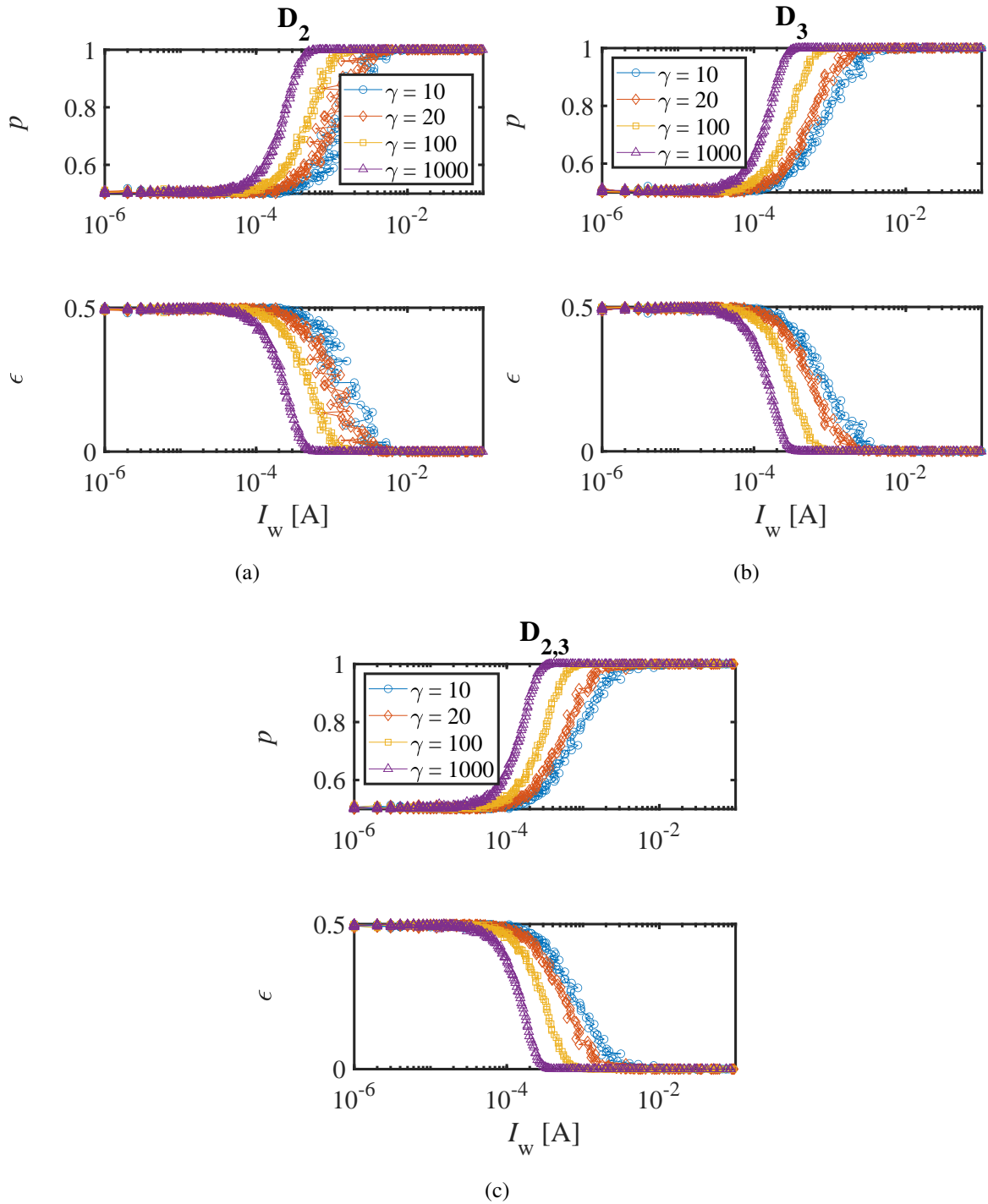

Figure 4.5: Eve's correct-bit-guessing probability $p$ (top) and Eve's bit error $\epsilon$ (bottom) with respect to the effective current $I_\text{w}$ for: $D_2$ (a), $D_3$ (b), and $D_{2,3}$ at $\gamma = 10$ (blue), $\gamma = 20$ (orange), $\gamma = 100$ (yellow), and $\gamma = 1000$ (purple). As $\gamma$ and $I_\text{w}$ (driven by the effective temperature) decrease, $p$ approaches perfect security.

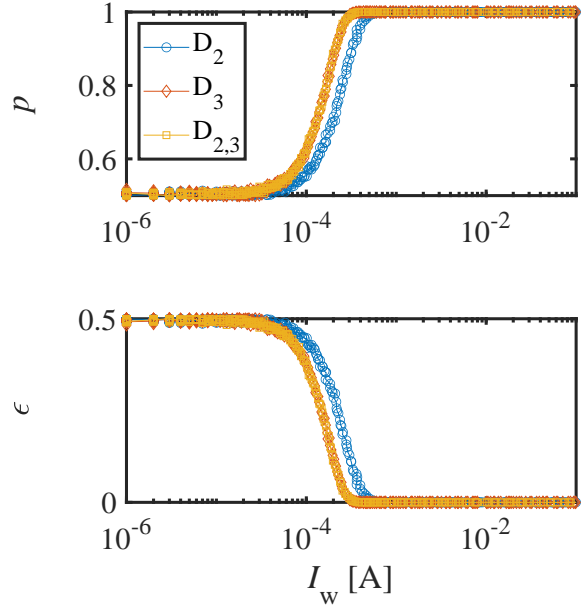

Figure 4.6: Eve's correct-bit-guessing probability $p$ (top) and Eve's bit error $\epsilon$ (bottom) with respect to the effective current $I_\text{w}$ at $\gamma = 1000$ for $D_2$, $D_3$, and $D_{2,3}$. $p$ increases and $\epsilon$ decreases as $I_\text{w}$ (driven by the effective temperature) increases. Convergence to perfect security happened at $D_2$ before $D_3$ and $D_{2,3}$.

### 4.3.1 Transition

This concludes the nonlinearity attack presented in this dissertation. As a take-home message, nonlinearity causes a nonzero power flow that leads to information leak. Now, we move onto our overall summary and conclusions.

# 5. SUMMARY AND CONCLUSIONS[1,2,4,5]

Secure key exchange protocols utilize random numbers, and compromised random numbers lead to information leak. So far, it had been unknown how Eve can utilize full or partial knowledge about the random number generators to attack the KLJN protocol. To demonstrate how compromised RNGs can be utilized by Eve, we have introduced four deterministic RNG attacks and four statistical RNG attacks on the KLJN scheme.

For the deterministic attacks, we showed that if Eve knows the root of both Alice's and Bob's RNGs, that is, when she exactly knows the random numbers, she can use Ohm's Law to crack the bit exchange. Eve can extract the bit very quickly, and she will learn the exchanged bit faster than Alice and Bob who know only their own random numbers. We showed that if Eve knows the seed of both Alice's and Bob's RNGs, that is, when she exactly knows the random numbers, she can crack the bit exchange even if her measurements have only one bit of resolution. The situation is similar to string verification in the noise-based logic systems. The cracking of the exchanged bit is exponentially fast; Eve can extract the bit within a fraction of the bit exchange period. Thus, Eve will learn the exchanged bit faster than Alice and Bob who know only their own random numbers.

We have also shown that if Eve knows the seed of only Alice's RNG, she can still crack the secure bit by using Ohm's Law or a process of elimination. However, she is required to utilize the whole bit exchange period.

---

[1]Part of this chapter is reprinted with permission from C. Chamon, S. Ferdous, and L. B. Kish, "Deterministic random number generator attack against the Kirchhoff-law-Johnson-noise secure key exchange protocol," *Fluctuation and Noise Letters*, vol. 20, no. 5, 2021. Copyright 2021 by World Scientific Publishing Company.

[2]Part of this chapter is reprinted with permission from C. Chamon, S. Ferdous, and L. B. Kish, "Statistical random number generator attack against the Kirchhoff-law-Johnson-noise secure key exchange protocol," *Fluctuation and Noise Letters*, accepted for publication, 2021. Copyright 2021 by World Scientific Publishing Company.

[4]Part of this chapter is reprinted with permission from C. Chamon, L. B. Kish, "Perspective–on the thermodynamics of perfect unconditional security," *Applied Physics Letters*, vol. 119, pp. 010501, 2021. Copyright 2021 by AIP Publishing.

[5]Part of this chapter is reprinted with permission from C. Chamon, S. Ferdous, and L. B. Kish, "Nonlinearity attack against the Kirchhoff-law-Johnson-noise secure key exchange protocol" *Fluctuation and Noise Letters*, in press, 2021. Copyright 2021 by World Scientific Publishing Company.

No statistical evaluation is needed, except for in the rarely-occurring event that Eve does not know which RNG belongs to which resistor, which will render in a waiting/verification (no-response) time that has a negligible effect on Eve's cracking scheme.

For the statistical attacks, we explored various situations how RNGs compromised by statistical knowledge can be utilized by Eve. We have introduced four new attacks against the KLJN scheme. The defense against these attacks is the usage of true random number generators (with proper tamper resistance) at Alice's and Bob's sides.

It is important to note that:

- To utilize these attacks, we implicitly used Kerckhoffs's principle/Shannon's maxim [2], which means Eve knows all the fine details of the protocol, including how the seeds are utilized and the RNG outputs timed.

- This exploration was done assuming an ideal KLJN scheme. Future work would involve a practical circuit implementation or a cable simulator and related delays and transients.

- Deterministic and statistical knowledge of the random number(s) by Eve is a strong security vulnerability. However, it is an illustrative way how such attacks can be developed.

We introduced a new passive attack that extracts information from the VMG-KLJN system and successfully compromises its security. On the other hand, the standard KLJN and the FCK1-VMG-KLJN schemes are immune against this attack because their net power flow is zero due to their thermal equilibrium feature. Our results prove that thermal equilibrium is essential for the perfect security of KLJN schemes, including their VMG-KLJN variations. Therefore, the Second Law of Thermodynamics is the fundamental component of the security of the VMG-KLJN system, too.

Nevertheless, we believe that, with careful design and proper compromises, the VMG-KLJN scheme [51] has strong potential for applications in chip technology even if its security level is reduced compared to the features of original KLJN protocol. It is a security level that is sufficient for many practical applications but is reduced due to the deviation from the original KLJN.

Its FCK1-VMG-KLN version [52] virtually eliminates the information leak, but then only three resistors can be freely chosen.

Finally, we introduced a new passive attack against the KLJN secure key exchange scheme when nonlinearity is present in the transfer function of the amplifier stage of noise generators. We demonstrated the effect of a 1% total distortion at the second order ($D_2$), third order ($D_3$), and a combination of the two orders ($D_{2,3}$) on the KLJN scheme.

We also demonstrated that, at a given nonlinear transfer characteristic, decreasing the effective voltage and, in this way reducing the nonlinearity, is a viable defense against the effect of nonlinearity in the KLJN scheme.

Our results showed that nonlinearity causes a notable power flow that leads to a significant information leak, so a careful design must be implemented such that the total distortion is kept at a minimum.

Alternatively, privacy amplification protocols [48, 50, 57, 99] can also be used. For example, as an active privacy amplification, Alice and Bob can also measure and compare the power flow and discard a proper fraction of high-risk bits [48, 50].